\documentclass[journal=jacsat,manuscript=article]{achemso}
\usepackage{CJKutf8}
\usepackage[version=3]{mhchem} 
\providecommand{\keyword}[1]{\textbf{Keywords:} #1}

\author{Hongbo,Li}
\affiliation[first unit]{Aerospace Information Research Institute, Chinese Academy of Sciences, Beijing 100094, China}
\alsoaffiliation[second unit]{Key Laboratory of Electromagnetic Radiation and Sensing Technology, Chinese Academy of Sciences, Beijing 100190, China }
\alsoaffiliation[third unit]{School of Electronic, Electrical and Communication Engineering, University of Chinese Academy of Sciences, Beijing 100049, China}
\alsoaffiliation[fourth unit]{GBA branch of Aerospace Information Research Institute, Chinese Academy of Sciences, Guangzhou 510700, China}
\alsoaffiliation[fifth unit]{Guangdong Provincial Key Laboratory of Terahertz Quantum Electromagnetics, Guangzhou 510700, China}
\author{Tianwu Wang}
\affiliation[first unit]{Aerospace Information Research Institute, Chinese Academy of Sciences, Beijing 100094, China}
\alsoaffiliation[second unit]{Key Laboratory of Electromagnetic Radiation and Sensing Technology, Chinese Academy of Sciences, Beijing 100190, China }
\alsoaffiliation[third unit]{School of Electronic, Electrical and Communication Engineering, University of Chinese Academy of Sciences, Beijing 100049, China}
\alsoaffiliation[fourth unit]{GBA branch of Aerospace Information Research Institute, Chinese Academy of Sciences, Guangzhou 510700, China}
\alsoaffiliation[fifth unit]{Guangdong Provincial Key Laboratory of Terahertz Quantum Electromagnetics, Guangzhou 510700, China}
\email{wangtw@aircas.ac.cn}
\author{Wenyin Wei}
\affiliation[fourth unit]{GBA branch of Aerospace Information Research Institute, Chinese Academy of Sciences, Guangzhou 510700, China}
\alsoaffiliation[fifth unit]{Guangdong Provincial Key Laboratory of Terahertz Quantum Electromagnetics, Guangzhou 510700, China}
\author{Kai Zhang}
\affiliation[fourth unit]{GBA branch of Aerospace Information Research Institute, Chinese Academy of Sciences, Guangzhou 510700, China}
\alsoaffiliation[fifth unit]{Guangdong Provincial Key Laboratory of Terahertz Quantum Electromagnetics, Guangzhou 510700, China}
\author{Jingyin Xu}
\affiliation[fourth unit]{GBA branch of Aerospace Information Research Institute, Chinese Academy of Sciences, Guangzhou 510700, China}
\alsoaffiliation[fifth unit]{Guangdong Provincial Key Laboratory of Terahertz Quantum Electromagnetics, Guangzhou 510700, China}
\author{Yirong Wu}
\affiliation[first unit]{Aerospace Information Research Institute, Chinese Academy of Sciences, Beijing 100094, China}
\alsoaffiliation[second unit]{Key Laboratory of Electromagnetic Radiation and Sensing Technology, Chinese Academy of Sciences, Beijing 100190, China }
\alsoaffiliation[third unit]{School of Electronic, Electrical and Communication Engineering, University of Chinese Academy of Sciences, Beijing 100049, China}
\alsoaffiliation[fourth unit]{GBA branch of Aerospace Information Research Institute, Chinese Academy of Sciences, Guangzhou 510700, China}
\alsoaffiliation[fifth unit]{Guangdong Provincial Key Laboratory of Terahertz Quantum Electromagnetics, Guangzhou 510700, China}
\author{Guangyou Fang}
\affiliation[first unit]{Aerospace Information Research Institute, Chinese Academy of Sciences, Beijing 100094, China}
\alsoaffiliation[second unit]{Key Laboratory of Electromagnetic Radiation and Sensing Technology, Chinese Academy of Sciences, Beijing 100190, China }
\alsoaffiliation[third unit]{School of Electronic, Electrical and Communication Engineering, University of Chinese Academy of Sciences, Beijing 100049, China}
\alsoaffiliation[fourth unit]{GBA branch of Aerospace Information Research Institute, Chinese Academy of Sciences, Guangzhou 510700, China}
\alsoaffiliation[fifth unit]{Guangdong Provincial Key Laboratory of Terahertz Quantum Electromagnetics, Guangzhou 510700, China}
\email{gyfang@mail.ie.ac.cn }
\title[An \textsf{achemso} demo]
  {Real-Space Sampling of Terahertz Waveforms Under Scanning Tunneling Microscope}

\abbreviations{IR,NMR,UV}

\begin{document}
\begin{CJK}{UTF8}{gbsn}
\begin{tocentry}
  \centering
\includegraphics[width=1\textwidth]{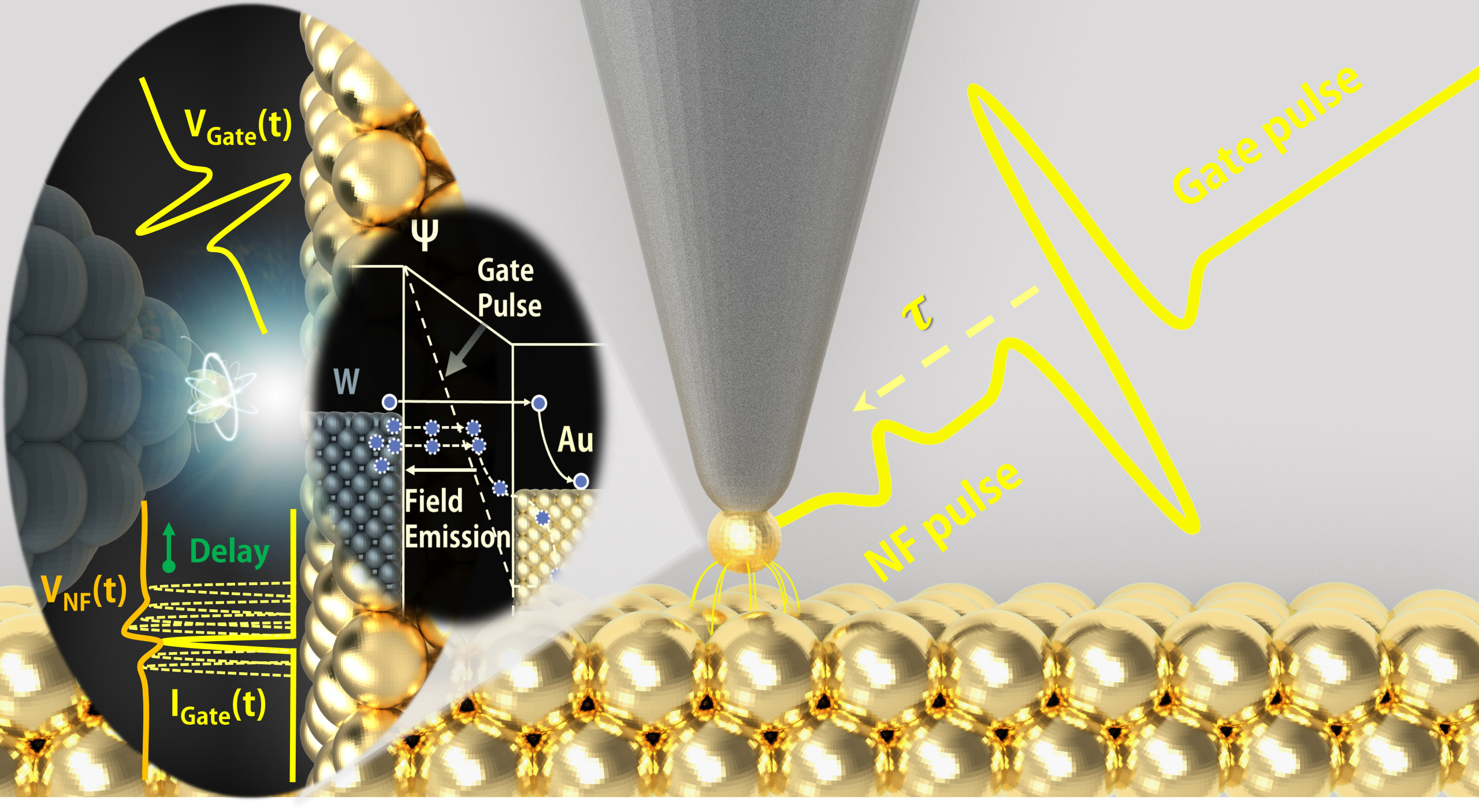}
\end{tocentry}

\begin{abstract}
Terahertz scanning tunneling microscopy (THz-STM) has emerged as a potent technique for probing ultrafastnanoscale dynamics with exceptional spatiotemporal precision,whereby the acquisition of THz near-field waveforms holds paramount significance. While substantial efforts have been dedicated to retrieving the waveform utilizing the photoemission current or a molecular sensor, these methods are challenged by intense thermal effects or complex sample preparations. In this study, we introduce a universal approach for real-time characterization of THz near-field waveforms within the tunnel junction,achieving subnanometer spatial resolution. Utilizing the gating mechanism intrinsic to the STM junction, coherent scanning of a gated strong THz pulse over a weak THz pulse is achieved,facilitating the direct measurement of the waveform. Notably, employing a custom-built carrier envelope phase (CEP) shifter, THz-CEP has been successfully characterized in the tunnel junction. Moreover, subnanometer spatial resolution of the THz-driven field emission current has been shown, underscoring the nanoscale resolving ability of our methodology. Ultimately, the potential application of this method for local THz time domain spectroscopy imaging has been demonstrated through point-to-point waveform sampling over the Au(111) surface. 
\end{abstract}
\keyword{\textbf{terahertz},\textbf{scanning tunneling microscopy}, \textbf{near-field}, \textbf{carrier-envelope-phase}, \textbf{ultrafast nanoscale},\textbf{nonlinear effect}}
\section{Introduction}
    The microscopic realm has undergone extensive exploration with the aid of high spatial resolution instrumentation\cite{RN1}, enabling profound insights into nanoscale materials\cite{RN2,RN3}. Among those advanced instruments, Scanning Tunneling Microscopy (STM) has made a revolutionary impact on our understanding of the microcosm due to its exceptional spatial resolution and versatile spectral capabilities, establishing it as a fundamental tool across diverse research domains\cite{RN4,RN5,RN6,RN7,RN8}. However, STM alone lacks the ultrafast temporal resolution required to study surface particle dynamics occurring on a sub-picosecond timescale\cite{RN9,RN10,RN11}, necessitating the development of time-resolved STM techniques\cite{RN12,RN13,RN14,RN15}. In these techniques, THz-STM has emerged as a promising choice with a cutting-edge spatiotemporal resolution\cite{RN16}, empowering investigations into ultrafast surface physics at sub-nanometers spatial scales and femtoseconds time scales. According to the Keldysh theory\cite{RN17,RN18}, the THz electric field is commonly regarded as a quasi-DC field applied to the tunnel junction, which varies within a THz cycle. Due to the intrinsic gating effect of the tunnel junction, THz current pulses with sub-picosecond durations can be achieved. These THz current signals, typically rectified through a current pre-amplifier\cite{RN16}, have been harnessed by numerous researchers to explore the potential applications of THz-STM in terms of imaging\cite{RN19,RN20,RN21,RN22,RN23}and pump-probe measurements\cite{RN16,RN22,RN23,RN24,RN25,RN26}. These achievements collectively underscore the versatility of THz-STM in probing ultrafast nanoscale dynamics and its vast potential across a spectrum of applications in surface physics and beyond.

    THz-STM offers a unique capability for researchers to investigate light-matter interactions at an unprecedented spatiotemporal scale, whereby the acquisition of THz near-field waveforms takes on critical significance. While the characterization of the far-field THz waveform can be achieved through a variety of techniques\cite{RN27,RN28,RN29}, its near-field measurement is still challenging. Substantial efforts have been dedicated to characterizing near-field THz waveforms by employing a metallic nanotip exposed to both near-infrared and THz pulses\cite{RN25,RN30,RN31,RN32}. However, this approach falls short of being truly in-situ due to the perturbation of the near-infrared radiation. More recently, there has been notable progress in achieving quantitative sampling of THz near-field waveforms within the tunnel junction, involving the superposition of two THz pulses through a single-molecule switch\cite{RN33}. Nevertheless, while the gating mechanism established using this molecular sensor is innovative, the complex sample fabrication process restricts its widespread applicability. Further research is urgently needed to generalize the superposition sampling approach by developing a more universally applicable gating mechanism, thus expanding the scope of its utility across diverse applications.
    
    In this article, we present a universal approach for the direct retrieval of THz near-field waveforms, leveraging an intrinsic gating mechanism provided by the Terahertz-driven field emission current. This approach employs a coherent modulation technique, whereby a weak THz pulse is superposed on a gated strong THz pulse, modulating its peak electric field strength. By intentionally chopping the weak THz pulse and subsequently demodulating the resulting current signal at various sweeping positions, the near-field waveform associated with the weak pulse is effectively retrieved across diverse surfaces. Evidently, the accuracy of this process heavily relies on the powerful gating of the strong pulse, a crucial aspect we will delve into later. Furthermore, by utilizing the recently developed THz Carrier-Envelope Phase (CEP) shifter\cite{RN34}, CEP variations in the THz near-field waveforms are successfully observed within the tunnel junction. Moreover, the sub-nanometer spatial resolution of our sampling methodology has been validated through meticulous point-to-point waveform sampling on a gold step and imaging across an atom cluster on the Au (111) surface with the waveform peak. Finally, an example of imaging with local THz time domain spectroscopy via this sampling approach is introduced. These findings significantly highlight our approach's capacity for real-time retrieval of sub-nanometer scale THz near-field waveforms, showcasing its generalized applicability and potential applications in surface physics for ultrafast nanoscale research.
\section{Methods and Results}
\subsection{Experimental Setup}

    The STM (Scanning Tunneling Microscope) system used in this study is a commercial ultrahigh vacuum (UHV) low-temperature STM with a base pressure of 7×10\textsuperscript{-11} mBar (Createc CT-1080). The STM tips employed in the experiments were fabricated by electrochemical etching of a 0.3 mm diameter tungsten wire. The preparation of a pristine Au (111) surface experienced a meticulous procedure, encompassing a combination of controlled annealing and ion etching. The STM experiments were conducted at a low temperature of 77 K using liquid nitrogen cooling and utilized sample bias with a preamplifier (Femto DLPCA-200), which has a gain of 10\textsuperscript{9}V/nA and a low-pass cut-off frequency of 1 kHz. The incident THz pulse was linearly polarized along the tip axis and was coupled to the STM tip at an incident angle of 30°. The THz pulse was chopped at 689 Hz, and the reference signal was sent to a lock-in current amplifier to demodulate the THz rectified current signal from the total tunnel current. The number of tunneling electrons driven by a THz pulse (\textit{N}) in the constant height mode was calculated using the equation $N=I/f$, where\textit{ I }represents the rectified current, \textit{f }is the repetition rate of the THz laser (which is 520.8 kHz in this experiment) and \textit{e} is the elementary charge. Precise control enabled us to achieve a tunneling current of less than 0.1 e/pulse. 

    In the optical setup (as depicted in \textbf{figure S1}), a ytterbium-doped potassium gadolinium tungstate (Yb:KGW) regenerative amplifier (Light Conversion CARBIDE) was used as the source of near-infrared laser pulses. This laser source provides pulses with specific parameters, including a repetition rate ranging from 1 kHz to 2 MHz, at a center wavelength of 1030±5 nm, and with a pulse duration of 199 fs. To generate terahertz (THz) radiation, an electro-optic crystal radiation method using a phase-matched LiNbO\textsubscript{3} crystal\cite{RN35,RN36} was adopted which produces THz pulses centered at 1 THz with a bandwidth of 2 THz. To measure the THz far-field waveform, electro-optical sampling (EOS) was performed using a 1mm ZnTe crystal\cite{RN37} and a probing light source with a pulse width of 199 fs. In all the TDFES measurements, humidity was maintained at around 15\% by purging nitrogen gas in a closed box covering all the THz beam route. 
\subsection{\textbf{THz-Driven Field Emission Sampling (TDFES) method}}
\begin{figure}
  \centering
\includegraphics[width=0.8\textwidth]{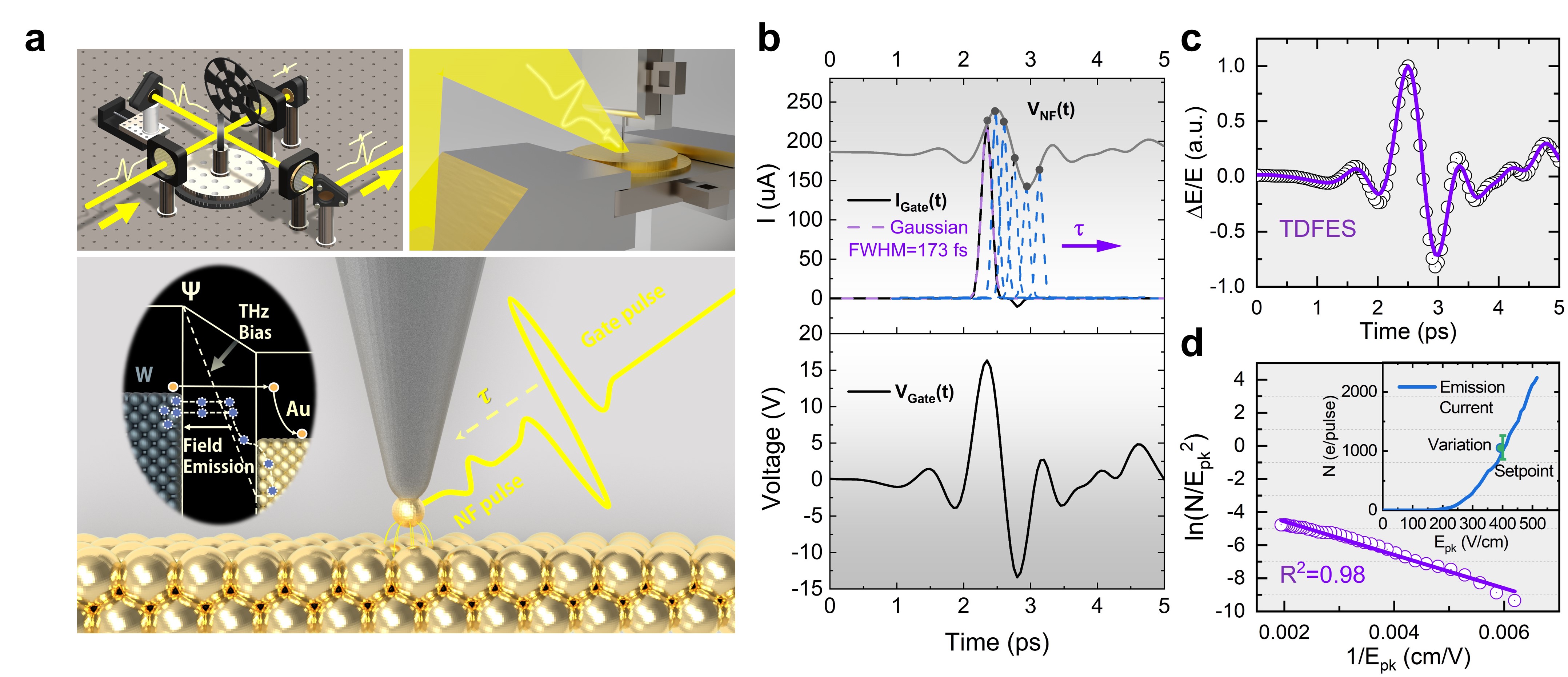}
  \caption{\textbf{Basic principle of the TDFES method} \textbf{a.} The THz beam comprises two pulses of varying amplitudes generated by a Michelson interferometer. The more robust THz pulse, referred to as the "Gate pulse," imparts a substantial local THz bias to the tunnel junction which significantly reduces the effective barrier width, resulting in the generation of clusters of emitted electrons (illustrated as blue dashed circles). In the energy band diagram, $\Phi$ denotes the vacuum energy level, the solid line represents the tunneling process driven by the static DC bias, while the dashed line with blue circles portrays the THz-driven field emission current. \textbf{b. }The current-tip height (I-Z) curve obtained at various DC bias levels and distinct THz illuminations. The THz strength is denoted by the maximum electric field peak intensity ($E_{pk}$) at the tip (see the Supporting Information). \textbf{c.} The average barrier height (ABH) retrieved from the curves presented in (b). \textbf{d. }The purple curve depicts the number of electrons tunneling per THz pulse (N) against THz current peak intensity $I(E_{pk})$ predicted by the Simmons model (see the Supporting Information), with a noticeable linear fit. The blue curve portrays the  \textit{$ln(N/E_{pk}^2))\sim 1/E_{pk}$} curve based on the experimental \textit{$N\sim E_{pk}$} data which is fitted to the Fowler-Nordheim theory. \textbf{e} The illustration of the sampling process in the TDFES where the stronger THz pulse is gated within the junction (the bottom chart), producing a narrowed current pulse with a Gaussian FWHM of 173 femtoseconds ($E_{pk}$=+300 V/cm), which is further modulated by the NF pulse (the middle chart). The top chart displays the experimental TDFES result, fitted using the Simmons model. The waveform shape was extracted from the experimental TDFES result measured at constant height mode under STM THz bias 3 V 20 pA. During the measurement, the DC bias was set to 3V, the Gate THz maximum $E_{pk}$ was +463.7 V/cm and the relative maximum $E_{pk}$ of the two pulses ($\eta = E_{NFpk}/E_{Gatepk}$) was 0.18.}
  \label{fgr:example}
\end{figure}

    As shown in Fig. 1a, the incident THz beam incorporates a pair of coherent pulses with distinct intensities from a classical Michelson interferometer. The two pulses are referred to as the "Gate pulse" and the "near-field (NF) pulse" respectively, in accordance with the notion in the previous work\cite{RN33}. The stronger THz pulse (denoted as the Gate pulse) exerts a local THz bias of a dozen volts\cite{RN20} onto the tunnel junction, which greatly lowers the barrier (indicated by a dashed line) and initiates the emission of clusters of electrons. In Fig. 1b, the current-tip height (I-Z) curve was recorded under various DC bias voltages and distinct THz illumination strengths in constant current mode. Simultaneously, the average barrier height (ABH) of the junction was calculated accordingly\cite{RN38}. Fig. 1c portrays the noteworthy reduction in the ABH of the tunnel junction under increasing DC bias voltages up to 9 V, with the ABH under THz illumination conspicuously lower than the value measured under DC bias. This observation strongly implies that the tunneling current induced by the THz electric field predominantly originates from field emission, given the significantly lower ABH under THz illumination compared to the work function of the tip or the sample\cite{RN20}.Additional evidence of the field emission essence for the THz current is provided by fitting the number of electrons tunneling per THz pulse (N) to the Fowler-Norheim (FN) theory\cite{RN39}. Here N is calculated using the formula: 
    \begin{equation}
N=1/e\int_0^T I_{THz}(t)dt, T=1/f_r \label{1}
\end{equation}
    where $f_r$ is the repetition rate of the THz pulses, $I_{THz}$ (t) is the transient THz current, and e is the elementary charge. Fig. 1d presents the prediction of N under varying THz maximum $E_{pk}$ using the Simmons model (refer to Supporting Information), revealing a linear relationship with $I(E_{pk})$. This not only underscores the predominant contribution of the THz maximum $E_{pk}$ in the total THz current but also validates the correlation expressed by the $N\sim E_{pk}$ curve. By fitting the $N\sim E_{pk}$ curve to the FN theory, the THz-induced tunneling current could be unequivocally identified as originating from THz-driven field emission\cite{RN20}. In FN theory, the tunneling probability of electrons under a DC electric field can be described as follows:
    \begin{align}
    &I=\frac{e^3}{2\pi h} \frac{\mu ^{\frac{1}{2}}}{(\chi+\mu)\chi^{\frac{1}{2}}} F^2 e^{-\frac{4\kappa\chi^{\frac{3}{2}}}{3eF}} \tag{2a}\\
    &\chi=C-\mu\tag{2b}\\
    &\kappa= \frac{\pi}{h} \sqrt{8m}\tag{2c}
    \end{align}
    Where I represents the tunneling current, C is the barrier height of the tunneling gap, μ is the thermodynamic partial potential of an electron (significantly lower than C), e is the elementary charge, h is the Plank constant, m denotes the mass of the electron, and F signifies the electric field strength. Considering the similarity of $N\sim E_{pk}$ with $I\sim F$, it is intuitive to fit the $N\sim E_{pk}$ curve in the form of  \textit{ln}$(N/(E_{pk}^2 ))\sim 1/E_{pk}$ as seen from Equation (2a), (2b) and (2c). As illustrated in Fig. 1d, the fitting results of  \textit{ln}$(N/(E_{pk}^2 ))\sim 1/E_{pk}$ exhibit a good linear fitting with a slope of $k= -1.01×10^3$. Here, the unit of $E_{pk}$ is V/cm, and $\chi$ closely approximates the barrier height C, which is around 4.6 eV retrieved from Fig. 1c. Considering the exponential part in formula (2a), the field amplification coefficient could be calculated if we assume F as the amplified local electrical field of the THz maximum $E_{pk}$, which is straightforward since THz maximum $E_{pk}$ dominates the total THz current. Therefore, the amplification coefficient $\beta$ for THz maximum $E_{pk}$ at the tip (V/cm) could be calculated as:
    \begin{align}
    &\beta = -\frac{4\kappa\chi^{3/2}}{300ek} \tag{3}
    \end{align}
    The result is approximately $2.2×10^5$, a value consistent with a previous result\cite{RN20}. This robust linear fitting substantiates that tunneling electrons primarily originate from THz-driven field emission\cite{RN20}. \\
    Instead of fabricating a molecule sensor\cite{RN33}, we simply employ the THz-driven field emission current to introduce a potent gating, enabling a more versatile and in-situ characterization of THz near-field waveforms. However, in this THz-driven field emission sampling (TDFES) scenario, it is imperative to assess the validity of the superposition sampling strategy because a weak gating in the STM junction might broaden the pulse width of the main peak and increase the intensity of the secondary peaks in the Gate current pulse, thus introducing inaccuracies. In order to comprehensively elucidate the sampling process inherent to the TDFES method, a current model was formulated based on the principles outlined by Simmons\cite{RN16,RN40,RN41,RN42}(see the Supporting Information). Experimental data from the measurement on the Au (111) surface with a tungsten tip were collected for model fitting. 
    As illustrated in Fig. 1e, the THz pulse could raise a local voltage of over 10 volts inducing a substantial field emission current reaching a level of tens of microamperes. This THz-induced field emission current is subsequently gated by the nonlinear current-voltage characteristics inherent to the STM junction, resulting in a pulse width at the 100-femtoseconds level. For instance, the THz current pulse, with a maximum electric field peak ($E_{pk}$) measuring +300 V/cm, is fitted with a Gaussian function, yielding a full width at half maximum (FWHM) of 173 femtoseconds. Additional results, presented in Fig. S2, demonstrate a linear increase in FWHM with THz maximum $E_{pk}$. However, it is noteworthy that even with THz maximum $E_{pk}$ as high as +700 V/cm, the resulting pulse width remains consistently small, measuring less than 200 femtoseconds. In addition, the bandwidth of the THz pulse emitted from the LiNbO3 crystal\cite{RN36} is 2 THz (see Fig. S1), which is further narrowed by the tip antenna effect\cite{RN19,RN33,RN43}, rendering the 200-fs-width gate pulse sufficient for sampling. Therefore, despite the involvement of thousands of electrons in tunneling within the junction, the gating effect remains sufficiently robust to ensure the effectiveness of the superposition sampling. After being gated by the tunnel junction, the current signal of the Gate pulse is subsequently modulated by the NF pulse. In this modulation process, an additional voltage is applied to the tunnel junction, varying with the NF electric field waveform. By adjusting the time delay between the two pulses, the near-field waveform of the weaker pulse can be extracted from the THz current.\\ 
    To simulate this process, the experimental TDFES result was employed as the target waveform. After normalization, it was multiplied by the maximum $E_{pk}$ of +300 V/cm to represent the Gate pulse. Subsequently, this Gate pulse was further multiplied by a factor of 0.05 to represent the NF pulse. The addition of the Gate and NF pulses was then fed into the current model, and the resultant overall current response was recorded while systematically varying the time delays of the two pulses. Following the subtraction of a DC component, the total current was normalized to represent the fitting waveform. As a result, we are able to reasonably reproduce the target TDFES waveform using our current model, as exemplified in Fig. 1e. 

\subsection{Validation and variable analysis of the TDFES method}
\begin{figure}
  \centering
\includegraphics[width=0.8\textwidth]{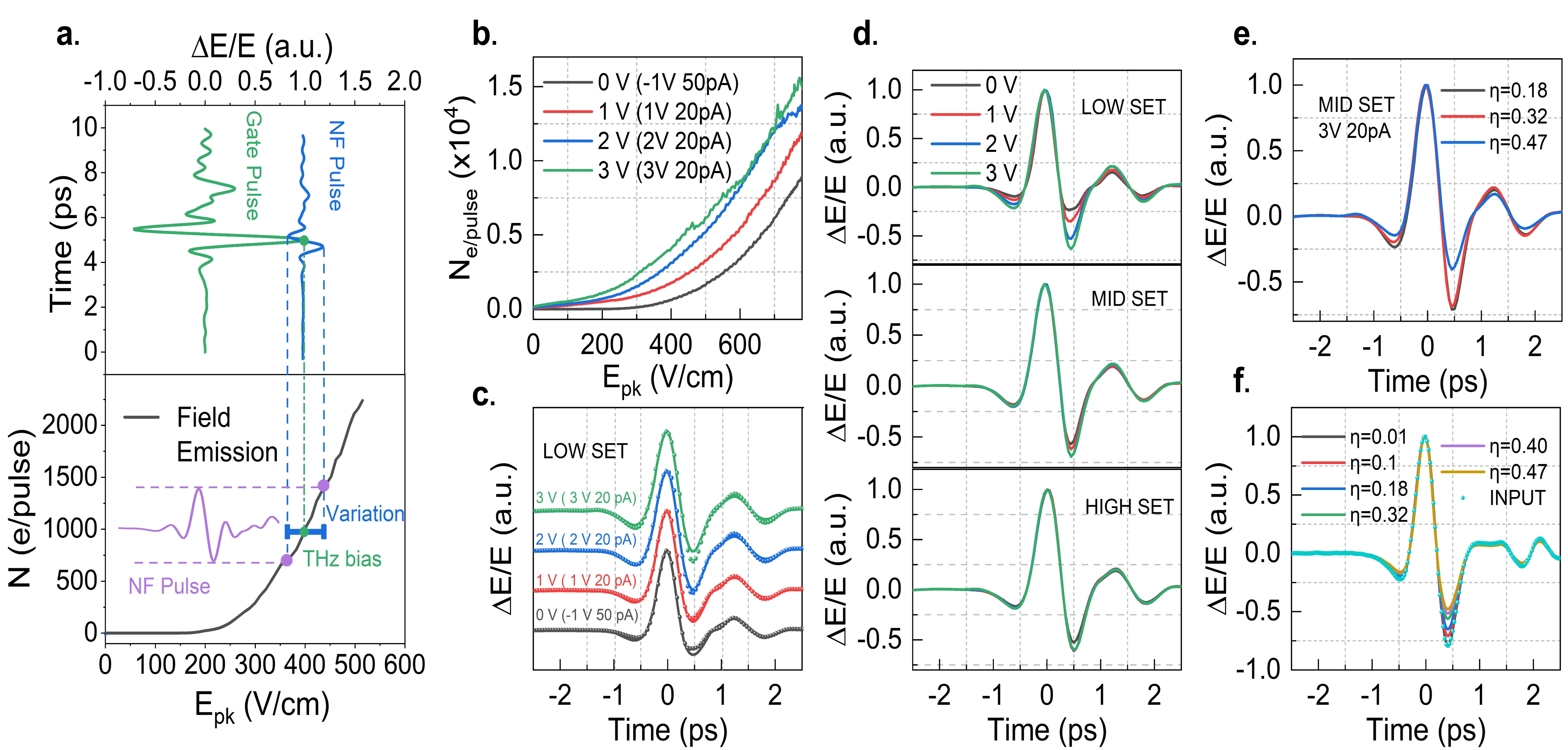}
  \caption{\textbf{Validation and variable analysis of the TDFES method} \textbf{a.} The illustration of the different roles played by the Gate pulse and the NF pulse on the $N\sim E_{pk}$ curve. The curve was measured at constant height mode at STM THz bias of 1 V 100 pA (without THz illumination), with the DC bias set to 0 V. \textbf{b} The $N\sim E_{pk}$ curves measured at four tip heights. 0V (-1V 50pA) represents for operating at constant height mode under STM THz bias -1 V 50 pA (without THz illumination), with the DC bias set to 0 V. Other conditions have analogous implications. The bias was introduced to enhance the local quasi-linearity of the $N\sim E_{pk}$ curve. \textbf{c}. Experimental TDFES results measured at a Gate pulse $E_{pk}$ of +358.9 V/cm, denoted by “THz bias”, with a relative maximum $E_{pk}$ of 0.32 defined by $\eta =(|E_{NFpk}|)/(|E_{Gatepk}|)$, under the four tip height settings in (b). A simulation based on the four $N\sim E_{pk}$ curves was carried out to fit the results. \textbf{d}. Experimental TDFES results measured with $\eta$ of 0.32 under the four tip height settings in (b), while varying the THz bias to +358.9 V/cm (top), +463.7 V/cm (middle), and +641.5 V/cm (bottom). \textbf{e}. Experimental TDFES results acquired at a THz bias of +463.7 V/cm with the tip height set at 3 V, 20 pA while changing $\eta$ of the two pulses. \textbf{f}. A simulation resembling (c) was conducted to explore the impact of varying $\eta$. (All the measurements used the same tip).}
  \label{fgr:example2}
\end{figure}
The linear fitting of the $N~I(E_{pk})$ curve, the fitting of $N\sim E_{pk}$ to the Fowler-Norheim theory, and the analysis of the secondary THz current peaks in the Supporting Information collectively support the conclusion that the contribution of the THz positive maximum electric field peak dominates the total transient Gate current pulse in TDFES. Consequently, a simplified explanation of the TDFES sampling process based on the $N\sim E_{pk}$ curve is proposed. As illustrated in Fig. 2a, the Gate pulse functions as a THz bias on the $N\sim E_{pk}$ curve, while the NF pulse introduces variations to this bias. Under the assumption of local linearity, the near-field waveform of the NF pulse could be retrieved from the variation of N. Mathematically, with the introduction of a small modulated variation to the THz bias, the Taylor series expansion of $N\sim E_{pk}$ yields:

\begin{equation}
N(E_{Gatepk}+\Delta E_{NF}\cdot h(t)=A_0+A_1\cdot \Delta E_{NF}\cdot h(t)+o^2(\Delta E_{NF}\cdot h(t))\tag{4}
\end{equation}	
where N represents the number of electrons tunneling per THz pulse, $E_{Gatepk}$ is the positive maximum electric field peak of the Gate pulse, $\Delta E_{NF}$ is the electric field of the NF pulse, $A_0$ is the N value at $E_{Gatepk}$, $A_1$ is the first derivative value of N at $E_{Gatepk}$, h(t) is the modulation function resembling a square wave in chopping modulation, and the term $o^2 (\Delta E_{NF}\cdot h(t))$ accounts for higher-order series components. As indicated by Equation (4), if $\Delta E_{NF}<<E_{Gatepk}$ is satisfied, the higher-order series terms become negligible. Consequently, the desired signal $\Delta E_{NF}$ can be easily extracted from the linear term $A_1\cdot \Delta E_{NF}\cdot h(t)$, which captures the weak THz waveform. As shown in the Taylor expansion, the accuracy of retrieving $\Delta E_{NF}$ from Equation (4) is highly reliant on the local quasi-linearity of the $N\sim E_{pk}$ curve.\\
    In the interpretation grounded on the \textit{N$\sim$E\textsubscript{pk}} curves, the TDFES method requires empirical validation to establish the soundness of its sampling logic, while the factors linked to the local quasi-linearity of the curve require further investigation. To experimentally clarify the method, a simulation was conducted to match the distorted TDFES waveforms presented in \textbf{figure 2b}. In detail, initially, the THz waveform shape was supposed to be identical at various tip heights, supported by experimental results in \textbf{figure S5} and evidence from previous studies\cite{RN31,RN32,RN33}. This waveform shape was selected from the TDFES result obtained at a THz bias of +463.7 V/cm under a tip height setting of 3 V 20 pA with a relative intensity of 0.18, utilizing the same tip. Next, the selected waveform was normalized and then multiplied by a THz bias of 358.9 V/cm along with the relative intensity of 0.32. This product was denoted as the variation and added to the THz bias together as input, which was subsequently computed using the \textit{N$\sim$E\textsubscript{pk}} curves in \textbf{figure 2a}. Finally, after subtracting the THz bias component, the calculated result was normalized to serve as the fitting curve and compared with the experimental curves. As depicted in \textbf{figure 2b}, the fitting results closely match the experimental ones, providing qualitative support for the modulation concept underpinning the TDFES method. As an additional validation measure, the transfer function of the TDFES spectrum relative to the far-field result was calculated. As shown in \textbf{figure S6}, these transfer functions comply with the $\sim$1/f law given by the tip antenna effect\cite{RN43} and are consistent with prior results\cite{RN32,RN33}, further reinforcing the method's validity.

    In the quest to comprehend the local quasi-linearity of the \textit{N$\sim$E\textsubscript{pk}} curve, a detailed investigation was conducted into three pivotal factors, including the curve itself, the THz bias on the curve, and the relative intensity of the two pulses. To explore the first two issues, TDFES experiments were carried out at three distinct THz bias and under four varied tip height settings associated with different \textit{N$\sim$E\textsubscript{pk}} curves. \textbf{figure 2c} shows the retrieved THz waveforms at three THz bias of +358.9 V/cm, +463.7 V/cm, and +641.5 V/cm, respectively, listed from top to bottom, while varying the tip height settings at each THz bias. As a result, the TDFES waveforms exhibit considerable variations when the THz bias is low (e.g., +358.9 V/cm) due to the poor local quasi-linearity. In contrast, the increase of the THz bias derives a converged waveform, signifying an improvement in the local quasi-linearity. This also implies the insensitivity of the near-field waveform to the changes in tip heights. In addition, at each THz bias, the TDFES waveform approaches the converged result with an elevated bias voltage, emphasizing the beneficial impact of higher bias voltage in enhancing local quasi-linearity.

    For the relative intensity$\eta$, three distinct values were checked experimentally at a THz bias of +463.7 V/cm under a tip height setting of 3 V, 20 pA. As illustrated in \textbf{figure 2d}, the measured waveforms show noticeable distortions as $\eta$ increases, particularly for the negative part of the waveform, whilst these distortions become much smaller at lower $\eta$ values. To further explore this, the simulation conducted for \textbf{figure 2b} was applied again with a THz bias of +463.7 V/cm and a tip height setting of 3 V, 20 pA, while varying $\eta$. As depicted in \textbf{figure 2e}, an increase in $\eta$ leads to discrepancies between the fitting waveforms and the input, aligning with the experimental observations in \textbf{figure 2d}. This phenomenon is attributed to the progressive intensification of a weak pulse, leading to the disruption of the previously established quasi-linear relationship in the \textit{N$\sim$E\textsubscript{pk}} curve. Consequently, it can be concluded that the TDFES method remains valid only when the relative intensity is sufficiently small, typically falling below 0.1, depending on the bias voltage and the THz bias. 

    In summary, the TDFES method has been successfully demonstrated and evaluated for its effectiveness in retrieving the near-field THz waveform within the tunneling junction. Nevertheless, its reliability is contingent on the local quasi-linearity of the\textit{ N$\sim$E\textsubscript{pk}} curve, which is influenced by several factors, including tip height settings, THz bias, and the relative intensity. To ensure accurate measurements, it is advisable to utilize high bias voltages, high THz bias, and maintain a small $\eta$. Furthermore, it's crucial to highlight that interference from THz current generated by other peaks of the gate pulse can pose challenges during scanning. As detailed in Fig. S2, an overly large ratio of the negative maximum $E_{pk}$ to positive maximum $E_{pk}$ in the Gate pulse shape, denoted as $\gamma =(|E_{pk-}|)/(|E_{pk+}|)$, may lead to inaccuracies in TDFES measurements. Consequently, a highly asymmetric pulse shape is anticipated. Despite potential interference from secondary current peaks, the simulation in Fig. S2 reveals that the similarity between the target waveform and the simulated waveform remains high under robust local linearity, even with a high $\gamma$ value of 0.8. In addition, the reduction of $\gamma$ from 0.8 to 0.2 only results in a minor decline by less than 1º in the inner product angle of the two Euclidean waveform vectors, suggesting that the $\gamma$ value in our experiment (0.8) is sufficiently low for accurate TDFES measurements under robust local linearity. Additionally, the simulation in Fig. S2 indicates that an increase in the THz bias significantly amplifies the influence of secondary current peaks, establishing an upper limit on the THz bias.

\subsection{CEP-varied THz waveform measurement }

    To further validate the effectiveness of the TDFES method, a homemade THz CEP shifter was employed to manipulate the weak THz pulse\cite{RN34}. Fig. 3a illustrates the modified experimental setup that allows unidirectional transmission of the THz beam, a requirement for the application of the CEP device. As outlined in the TDFES method, the near-field waveform is extracted from the weak THz pulse that modulates the peak intensity of the Gate pulse. Consequently, only the weak pulse passed through the CEP shifter, allowing for four different CEP shifts: $-\pi/2,0,\pi/2 and \pi$. The CEP device comprises a pair of metamaterial chips with intricately designed patterns determined by parameters $\alpha,\beta,\gamma,r_1, and r_2$ for phase shifting\cite{RN34}. To comprehensively demonstrate the efficacy of the CEP shifter and further verify the TDFES method, THz waveforms with varying CEPs were sampled using three methods including near-field TDFES, far-field EOS\cite{RN29} and near-field PES\cite{RN25} in the time domain. 
    
    In the photoemission sampling (PES) experiment, intensive 515 nm laser pulses were generated using a BBO crystal through a second harmonic generation process (SHG) with subsequent filtering to remove the 1030 nm component. The 515 nm laser beam was directed towards the STM tip, and the relative distance between the THz and laser beams could be adjusted using an electrical motion stage. Within the tunnel junction, these intense 515 nm laser pulses could induce a significant photoemission current, reaching up to 1 nA, with the gap size typically ranging from 1 to 10 μm. This current arose from a multiphoton absorption process occurring either at the tip or the sample surface, as depicted in Fig. S3. In the experiment, we opted to utilize the photoemission current emitted from the sample surface, as it offered a higher signal-to-noise ratio (SNR).

\begin{figure}
  \centering
\includegraphics[width=0.8\textwidth]{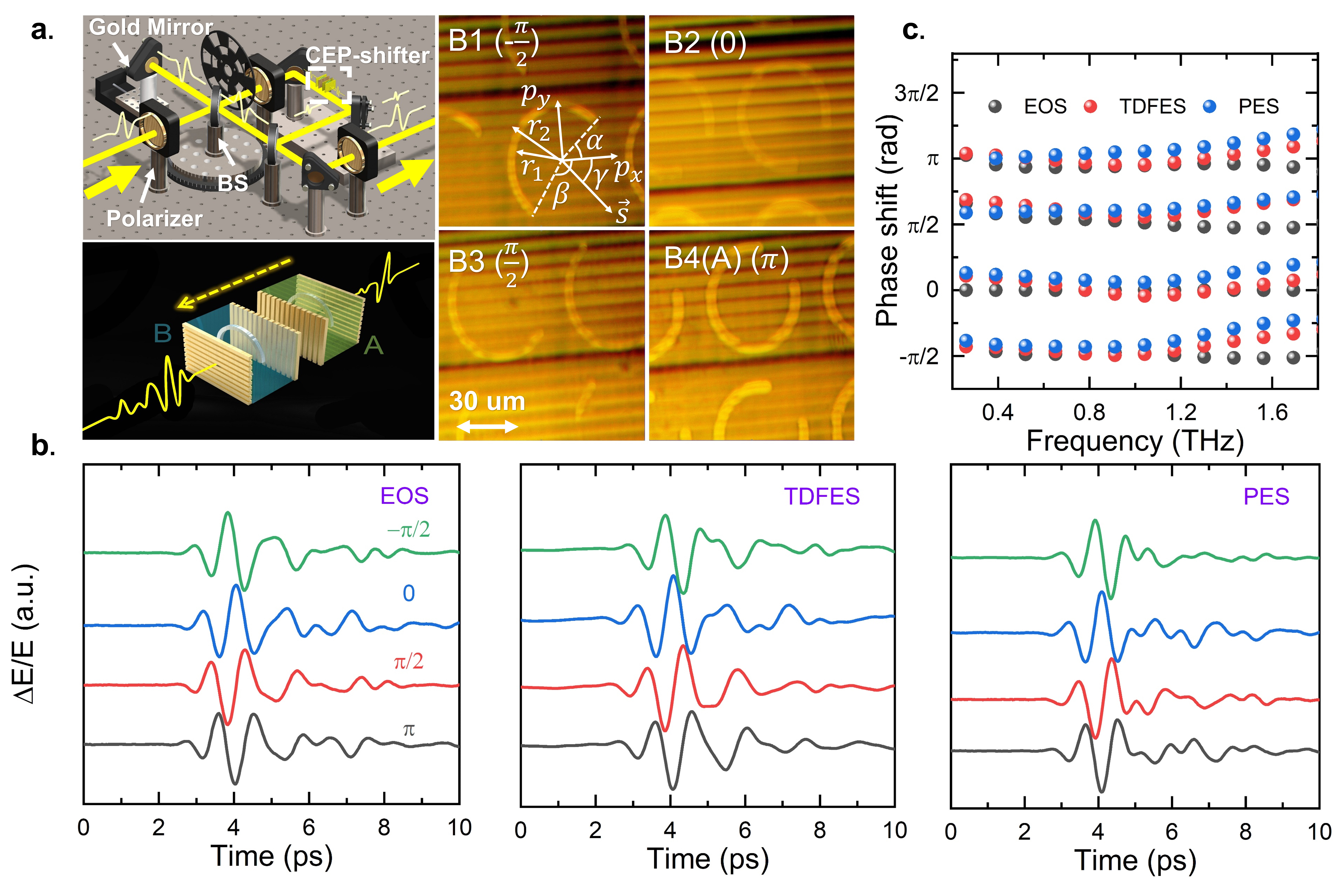}
  \caption{\textbf{Applying CEP shifter to obtain CEP-varied near-field THz waveforms a.} The working principle of the CEP device in which the Michelson interferometer is replaced by a new layout that allows unidirectional transmission of a single THz beam. The CEP shifter comprises two parts ($A \& B$) where A is fixed and B can select from four patterns to realize phase shifts of $-\pi/2,0,\pi/2  and  \pi$. Each pattern is designed with parameters $\alpha,\beta,\gamma,r_1, and r_2$ to realize different CEP changes. \textbf{b. }The waveforms measured using three different sampling methods, as depicted from left to right: EOS (far-field), TDFES, and PES. \textbf{c. }The CEP values calculated from all the waveforms obtained in \textbf{(b)}, utilizing their phases in the cross power spectral density (CPSD) with reference to the EOS result of zero CEP. In the TDFES measurements, the THz bias was 504.9 V/cm under tip height setting of 3 V, 20 pA (bias set to 3 V, constant height mode) with a relative intensity of 0.1. In the PES experiments, 515 nm laser pulse energy was 0.6 $\mu$J, the tip-metal gap was set at approximately 1 $\mu$m, and the PES current curve from \textbf{figure S3d }was used. (Note: All the experiments used the same tip). }
  \label{fgr:example3}
\end{figure}

    In\textbf{ figure 3b}, the THz waveforms obtained by different sampling methods all exhibit varying CEP features, including four phase shifts at $-\pi /2,0,\pi /2, \pi$, respectively, while maintaining a similar envelope. Importantly, to eliminate tip effects \cite{RN25,RN42,RN44,RN45} as indicated in \textbf{figure S5}, all measurements employed the same tip. A detailed comparison of the THz CEPs obtained through these sampling methods is presented in \textbf{figure 3c}. It reveals a slight CEP divergence among these methods, which varies with different CEP values and becomes more pronounced as the frequency increases. This divergence signifies the variation of the CEP in the vicinity of the STM tip which is unavailable in the EOS (far-field) results. Notably, the divergence observed in PES relative to the EOS result is more significant than that in TDFES. This distinction can be attributed to the different regions where the two methods are applied and the potential influence of the incident visible light pulse in PES, whose result is varied with the laser strenghth\cite{RN31}. Specifically, PES waveforms were measured using photoemission current from the sample side, which may be susceptible to thermal effects from the sample. Nevertheless, it has been demonstrated that relying solely on the sample photoemission current is sufficient for retrieving the near-field waveforms. This is contributed to the fact that the acquired PES results exhibit minor differences except for an inverted CEP when utilizing the tip emission current, as shown in \textbf{figure S3.} In conclusion, we have achieved relatively consistent waveforms across these methods with minor distinctions, underscoring the validity of the TDFES method and its capability of retrieving real-time local THz CEP features with higher accuracy.

\subsection{\textbf{Sub-nanometer scale TDFES measurements} }
\begin{figure}
  \centering
\includegraphics[width=0.8\textwidth]{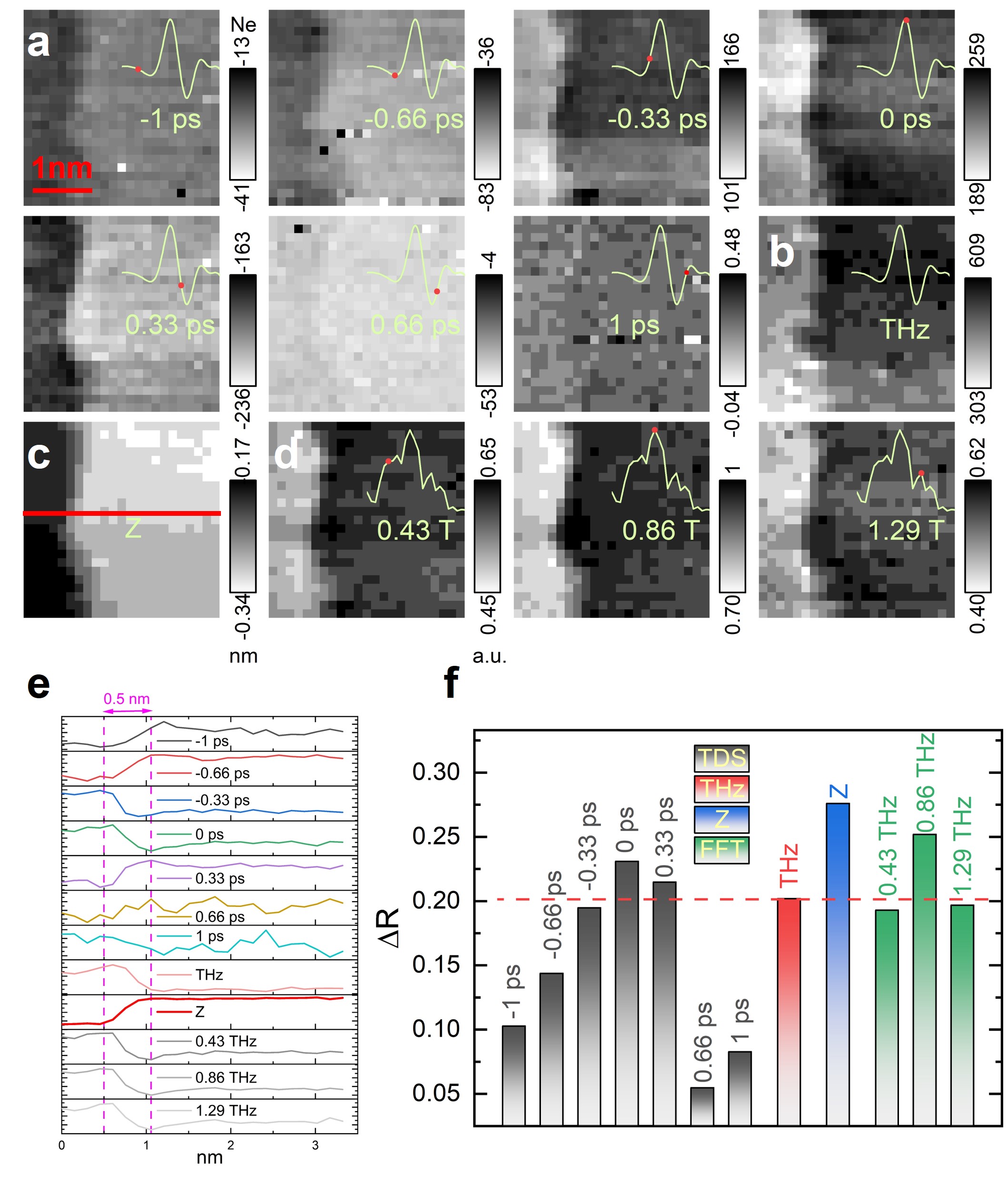}
  \caption{\textbf{Point-to-point sampling of THz-TDS on an Au (111) triatomic step} \textbf{a. } The STM topography data depicting a gold step with THz illumination on (constant current mode). STM setpoint was at 1 V 100 pA. \textbf{b}. Corresponding representation of the same area using THz rectified current at the identical height in (a), acquired with STM in constant height mode and a THz maximum $E_{pk}$ of +455 V/cm. \textbf{c}. TDEFS waveform recorded along the line marked by gray crosses on the gold step (top inset chart). The peak intensities of the waveforms were recorded. The tip height remained consistent with (a) during the measurement conducted in constant height mode, with a THz maximum $E_{pk}$ of +455 V/cm and $\eta$ set at 0.05. \textbf{d}. Large-scale surface images on Au (111), including a zoomed-in area featuring a cluster of Au atoms. STM setpoint remained consistent with (a) under constant current mode. \textbf{e}. Three types of images obtained at the same tip height as in (a) over the zoomed area in (d), with their line profiles displayed. Images were mapped at constant height mode by: DC current (DC bias was 1 V), THz waveform peak (DC bias was 0 V), and THz rectified current (DC bias was 0 V), from top to bottom. \textbf{f-h}. The same step as in (a) imaged through point-to-point sampling of THz-TDS signals: (f) waveform intensity mapping at temporal positions along the THz waveform (-1.6 ps~1.6 ps, time interval 0.2 ps); (g) Fourier-transformed spectrum intensity mapping at frequency positions ranging from 0 THz to 2.5 THz, with an interval of 0.3125 THz; (h) phase mapping at the same frequency positions as in (g). For all imaging experiments in (f)-(h), waveforms were measured under the same conditions: THz maximum $E_{pk}$ was +455 V/cm, $\eta$ was 0.05, and the tip height matched that in (a).}
  \label{fgr:example4}
\end{figure}
    To leverage Terahertz Time Domain Spectroscopy (THz-TDS) information at a local scale, the spatial resolution of TDFES for distinguishing waveform divergences within a confined area is elucidated. Notably, TDFES relies on field emission current for acquiring near-field THz waveforms within the tunnel junction, thus the lateral spatial resolution of the field emission current is investigated instead, which is sensitive to tip height variations (Fig. S5). Previous experiments using laser-excited THz radiation under STM conditions achieved sub-100 nm lateral resolution on a gold film atop an InAs substrate, indicating potential for local THz-TDS imaging. In this study, the spatial resolution is pushed to a sub-nanometer level using TDFES. 
    
Fig. 4a and Fig. 4b illustrate the imaging of a gold step on the Au (111) surface with topography information in constant current mode and THz rectified current in constant height mode. To determine spatial resolution, waveforms along a step line (Fig. 4a) were sampled point by point at constant height mode with a bias set to 0 V. Fig. 4c shows that due to varying tip height, TDFES waveform intensity fluctuates along the step with a spatial resolution of around 1.6 nm. However, normalized data suggests identical waveforms on both sides of the gold step. Better results are obtained by imaging an area with a cluster of Au atoms (Fig. 4d). The atom cluster displays sub-nanometer features imaged by DC current, TDFES waveform peak, and THz rectified current signals in constant height mode. Fig. 4e indicates that the atom cluster is well-recognized by the waveform peak intensity mapping, showing a Full Width at Half Maximum (FWHM) of 0.85 nm, confirming the ability of TDFES to detect waveform changes at a sub-nanometer scale.

Furthermore, Fig. 4f-Fig. 4h demonstrate using THz-TDS for imaging by mapping current values at different temporal positions along the waveform (Fig. 4f), obtaining THz spectral images by Fourier transforming time-domain waveforms (Fig. 4g) and retrieving phase information (Fig. 4h). In Fig. 4f and Fig. 4g, the THz-TDS images at diverse temporal and spectral positions exhibit varying image contrasts, with positions featuring higher intensity achieving superior recognition. At certain positions (e.g., 0 ps and 0.9 THz), nanometer-scale features of the gold step have been characterized. However, the phase images in Fig. 4h show no differences on either side of the gold step, indicating that the waveform shape is consistent on both sides of the step. Notably, TDFES responds a little differently to Au (111) and highly oriented pyrolytic graphite (HOPG) surfaces (Fig. S7), showcasing its versatility for applications across diverse surfaces and potential exploration of material properties within the tunnel junction. However, maintaining the mono-peak characteristic of the Gate current pulse is crucial for recognizing disparities in spectral profiles during imaging on different surfaces. In-depth investigations are warranted to ascertain the applicability of local THz-TDS information from TDFES for exploring material properties, especially in fabricating samples containing varying local THz-TDS information.

\subsection{Conclusion}
    In conclusion, the TDFES method has been successfully demonstrated to sample THz near-field waveforms in a tunnel junction with sub-nanometer spatial resolution. Its validity has been confirmed by simulation results and experimental comparations, where powerful gating of the Gate pulse is essential which favors highly asymmetric waveform shape with mono-peak THz current pulse. Rigorous validation and statistical analysis have highlighted the importance of local quasi-linearity in the field emission current curve for measuring real-space THz waveforms, which prefers high THz bias, high bias voltages and small $\eta$. In addition, a THz CEP device has been utilized to control the phase of the near-field waveform, which is directly characterized within the tunnel junction using the TDFES method. Furthermore, the sub-nanometer spatial resolution of the TDFES method has been demonstrated by point-to-point waveform sampling along a gold step and by imaging over a cluster of Au atoms based on the waveform peak intensity. The capability of TDFES for Terahertz Time Domain Spectroscopy (THz-TDS) imaging is exemplified in the local retrieval of THz-TDS information over an Au step in a point-by-point fashion.
    
In the far-field domain, THz-TDS techniques have become vital tools for material science\cite{RN46,RN47,RN48,RN49}, nevertheless, their spatial resolution is often limited to the micrometer scale. By overcoming the diffraction limit, scattering-type THz atomic force microscopy has pushed this resolution to sub-20 nm\cite{RN50,RN51,RN52} and THz-STM has further improved it to the atomic scale\cite{RN16,RN19,RN20,RN21,RN22,RN23}. However, it is still challenged to apply spectrum information in these near-field microscopies. Through our effort to build an in-situ sampling method, real-space THz-TDS has been directly obtained within the tunnel junction, expanding its utility in CEP identification, THz spectral imaging, and other potential applications. The work demonstrates the potential of the TDFES method to provide new opportunities for combining local THz-TDS with STM setups, opening fresh avenues of research in this field.

\begin{suppinfo}
The Supporting Information is available free of charge at \\https://pubs.acs.org/doi/10.1021/acsphotonics.3c01451. \\
Figuration of the optical setup, details of the samplepreparations, explanation of the Simmons model,illustration of the photoemission sampling, calibrationof the THz electric field strength, calibration of the THz-STM rectified current signal, and additional Figures S1−S7 (PDF)
\end{suppinfo}
\subsection{Corresponding Authors}
*E-mail: wangtw@aircas.ac.cn; gyfang@mail.ie.ac.cn.
\subsection{Author contributions}
G.-Y.F. and T.W. supervised the work. H.L. and T.W. built the THz-STM system. H.L. and W.W. prepared the samples and performed the measurements. H.L. analyzed the data with the help of T.W., K.Z. and J.X., H.L. and T.W. wrote the manuscript.
\subsection{Competing interests}
The authors declare no competing interests.
\begin{acknowledgement}
This work is financially supported by the National Natural Science Foundation of China (61988102,12274424), the National Key Research and Development Program of China(2022YFA1203500), Key Research and Development Program of Guangdong Province (2019B090917007), Science and Technology Planning Project of Guangdong Province (2019B090909011), Guangzhou basic and applied basic research Project (202102020053). 
\end{acknowledgement}
\bibliography{Manuscript}
\end{CJK}
\end{document}


\renewcommand*{\thefigure}{S\arabic{figure}}

\textbf{This file includes 15 pages containing:\\}
\textbf{Additional manuscript\\}
\textbf{Supplementary Figures S1 to S7\\}
\textbf{References\\} 
\section{Additional manuscript}
\subsection{Details of the optical setup}
    The experimental optical setup, as illustrated in figure S1a, is divided into three main components: the THz generation system (highlighted in red), the THz detection system (in blue), and the THz transmission and guiding system (in green). The generation and detection of THz radiation relied on a $LiN_bO_3$ crystal\cite{RN1,RN2} and the electro-optical sampling\cite{RN3} (EOS) method. Components in this setup included BS1 (a 2:8 beam splitter), lenses L3, L4, L1, and L2 with focal lengths of 150 mm, 50 mm, 150 mm, and 100 mm, respectively, and G1, a grating with 1600 lines/mm. Lenses L3 and L4 were used to adjust the spot size of the 1030 nm light on the $LiN_bO_3$ crystal surface to enhance THz emission, and these optics were mounted on high-precision motion stages for emission optimization. As seen in figure S1b and c, a typical THz pulse was generated and measured using a 1 mm ZnTe crystal, resulting in a waveform centered at 1 THz. 
    THz transmission was facilitated through off-axis parabolic mirrors and Au-coated reflection mirrors. A Michelson interferometer layout, employing a 1:1 THz beam splitter (BS2), was adopted, and wire grid polarizers (WG) were utilized for THz power adjustment. T1 represents a THz lens with a focal length of 50 mm. M1, M2, and M3 (enclosed in dashed boxes) are gold-coated mirrors placed on magnetic holders, where M1 can be rotated to form a Michelson interferometer or allow one THz beam to pass unidirectionally through the CEP device. Proper combinations of M1 and M2 can be used to invert the phase of the THz pulse. M3 was flexible for guiding the THz beam into the STM or measuring THz far-field electric fields. The guiding system accommodated a 515 nm green light beam generated by a BBO crystal and included a perforated parabolic mirror that allowed the combination of the THz beam with the green light. The 515 nm laser pulse served not only for guiding the THz beam but also for photon emission sampling experiments\cite{RN4}.
\begin{figure}
    \centering
    \includegraphics[width=0.8\textwidth]{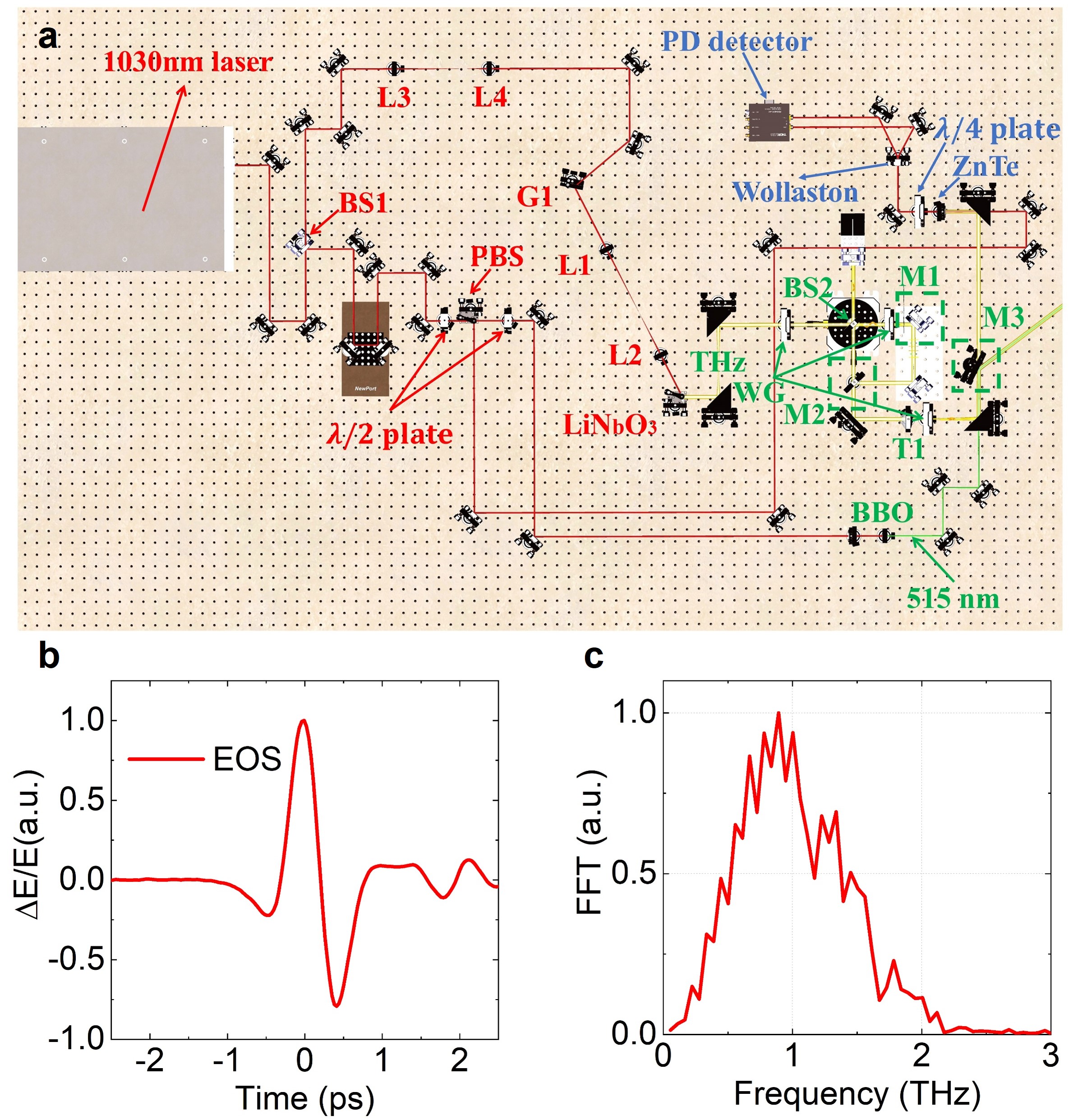}
    \caption{\textbf{Details of the Optical setup a.} Illustration of the optical setup \textbf{b.} The generated THz waveform measured using the EOS approach. \textbf{c.} The corresponding FFT-transformed result. }
    \label{fig:enter-label1}
\end{figure}
\subsection{\textbf{Sample Preparations}}
   The preparation of a pristine Au (111) surface involves a meticulous procedure, encompassing a combination of controlled annealing and ion etching. In the annealing process. the sample was gradually heated to 900 K, held at this temperature for 1 minute, then reduced to 800 K for 2 minutes, and finally allowed to cool to the ambient temperature of 300 K. Subsequently, the sample underwent argon ion etching, with cycles lasting 15 to 20 minutes and an etching current of 5 $\mu$A, ensuring the removal of impurities and contaminants. This two-step procedure was repeated for two times, ended with a final annealing at an ultrahigh vacuum of 10\textsuperscript{-10} mbar. For the highly oriented pyrolytic graphite (HOPG) surface, the surface layers were gently removed using adhesive tapes in an ambient air environment followed with an annealing procedure at 400 K for 10 minutes under ultrahigh vacuum conditions, ensuring a clean and well-prepared surface for experimental use.

\subsection{\textbf{\textbf{Quantify the current with the Simmons model}}}
    When coupled with the tungsten tip, THz pulses could induce an intense local electric field on the Au (111) surface, resulting in the emission of tens of electrons and the formation of a giant field emission current. This emission current pulse has a significant current peak, reaching tens of microamperes\cite{RN5} and a pulse width on the order of 100 femtoseconds. In the output of the current pre-amplifier, due to the limited bandwidth, the THz current was rectified through integration, calculated as follows\cite{RN6}:
\begin{equation}
    I_{rec}=\int I_{THz}(t)dt=\int I_{DC}(V_{THz}(t))dt \tag{1}
\end{equation}
Hereby, \textit{I\textsubscript{rec}} is the THz rectified current, \textit{I\textsubscript{THz}} is the THz current pulse, \textit{I\textsubscript{DC}} is the tunnel current driven by a DC bias, and \textit{V\textsubscript{THz}}is the THz induced voltage in the tunnel junction. \textit{I\textsubscript{DC}} was calculated from the static current model, which was obtained by fitting the \textit{I\textsubscript{DC}$\sim$V\textsubscript{DC}} curve without THz illumination\cite{RN7}. To obtain this model, the \textit{I\textsubscript{DC}$\sim$V\textsubscript{DC}} characteristic was assumed to follow the Simmons model\cite{RN6,RN8,RN9}, and the following formula was used for calculation\cite{RN8}: 
\begin{align}
        J_{DC}=\frac{2e^3(\frac{F}{\beta}^2)}{8\pi h\varphi_0} (exp(-\frac{4\pi\beta}{eF}m^{\frac{1}{2}}\varphi_0^{\frac{3}{2}})-(1+\frac{2eV_{DC}}{\varphi_0})exp(-\frac{4\pi\beta}{eF}m^{\frac{1}{2}}\varphi_0^{\frac{3}{2}}(1+\frac{2eV_{DC}}{\varphi_0})^{\frac{1}{2}}))
\tag{2}
\end{align}

where \textit{J\textsubscript{DC}} represents the current density, \textit{V\textsubscript{DC}} is the voltage applied on the gap, \textit{e }is the elementary charge, \textit{h} is the Planck constant, \textit{m} is the mass of an electron, $\varphi_0$ is the effective work function, and $\beta$ is 23/24 according to the Simmons model. \textit{F} denotes the local eletric field given by $F=\frac{V_{DC}}{S}$, where \textit{S} is the tunnel barrier width. To account for the asymmetric behavior of the \textit{I\textsubscript{DC}$\sim$V\textsubscript{DC}} curve and the THz-driven emission current curve (\textit{N$\sim$E\textsubscript{pk}}) observed in the measurements on the Au (111) surface with a tungsten tip, different effective work functions were assumed for the positive and negative curves. This asymmetry has been observed in the average barrier height (ABH) results retrieved from the static current-tip height (I-Z) curves, as shown in Fig. S2a. In addition, a 3D tip configuration\cite{RN5} was adopted for fitting. Thus, the total current was expressed as:

\begin{equation}
    I_{DC}=\int J_{DC}(S(x),V_{DC})2\pi xdx\tag{3}
\end{equation}
and the tip figuration was modeled as:

\begin{equation}
S(x)=
\begin{cases}
    Z+(r-\sqrt{r^2-x^2}),x\leq \sqrt{h(2r-h)}\\
    Z+h+\sqrt{R^2-h(2r-h)}-\sqrt{R^2-x^2},R\geq x\geq \sqrt{h(2r-h)}
\end{cases}
    \tag{4} 
\end{equation}
The parameters \textit{Z, r, h, R, x}, and \textit{S(x)} are illustrated in\textbf{ figure S2a}, where \textit{Z} represents the tip-sample minim distance, \textit{r} denotes the radius of the attached atom, \textit{h} describes the degree of the adhesion, \textit{R} is the radius of the needle,\textit{ x} denotes the distance away from the center and \textit{S(x) }denotes the barrier height at position \textit{x}. The current contribution is neglected beyond the radius of the needle \textit{R}. To calculate the rectified current of the \textit{V\textsubscript{THz}(t)}, TDFES experiments were performed on Au (111) to obtain the THz near-field waveform. As shown in Fig. S2c and S2d, our model reveals a compelling match with the recorded $I_{DC}\sim V_{DC}$ curve and the $N\sim E_{pk}$ curve, particularly in replicating the asymmetry observed in the $N\sim E_{pk}$ curve. This asymmetry, where the positive maximum peak of the THz electric field waveform is notably amplified compared to the negative peak within the tunnel junction, results in a positive rectified current even when confronted with a negative THz maximum $E_{pk}$. 

Employing this model, we delved into the intricacies of secondary peaks in the THz transient current during TDFES sampling. Our investigation, portrayed in Fig. S2e, elucidates that at higher humidity, diverse peaks in the THz electric field waveform contribute to the THz transient current, intricate due to water absorption. However, as humidity diminishes, the positive and negative maximum peaks dominate, owing to robust gating in the tunnel junction. Consequently, we concentrated solely on the influence of the negative maximum electric field peak in the TDFES experiment. This investigation was conducted with a Gate pulse shape, parameterized by $\gamma_0$, shaping the THz waveform as: 
\begin{equation}
V_{THz}(t)=
\begin{cases}
    1.268\cdot exp(-3(t-5)^2)sin(1.6\pi(t-5)),0\leq t\leq 5\\
    1.268\cdot exp(-\gamma_0(t-5)^2)sin(1.6\pi(t-5)),5\leq t\leq 10
\end{cases}
    \tag{5} 
\end{equation}
Here, parameter$\gamma_0$ is adjustable to vary the THz waveform shape and the constant parameters are designed to normalize the maximum value of $V_{THz}(t)$ to 1. As shown in Fig. S2f, nine distinctive Gate pulse shapes, each producing a maximum $E_{pk}$ of +300 V/cm, were modeled, and their ensuing current responses were predicted. The ratio of the negative maximum $E_{pk}$ of the Gate pulse to the positive maximum $E_{pk}$ is denoted as $\gamma=(|E_{pk-} |)/(|E_{pk+}|)$. The NF pulse, derived from experimental TDFES data, was incorporated by scaling it with the maximum $E_{pk}$ of +300 V/cm and the relative maximum $E_{pk}$ ($\eta=E_{NFpk}/E_{Gatepk} =0.05$). This product was then superimposed with the Gate THz electric field and subjected to the current model, systematically sweeping the NF pulse with the Gate pulse. After eliminating a DC component, the output current was normalized to its maximum absolute value, serving as the fitting result. The congruence between the fitting waveform and the target waveform was quantified through the inner product angle in Euclidean space.  
\begin{equation}
Inner product angle=arccos(\frac{\Sigma_nF_1(n)\cdot F_2(n)}{\sqrt{\Sigma_mF_1(m)^2\cdot\Sigma_mF_2(m)^2}})
    \tag{6} 
\end{equation}
Here, $F_1, F_2$ denote the waveform vector of the two pulses respectively, which are normalized by dividing the maximum absolute value. The smaller the angle is, the better the fitting results are. 
As shown in Fig. S2g, the NF pulse shape is adeptly reproduced with judiciously selected values of $\gamma$ and $\eta$. Nevertheless, Fig. S2h underscores the necessity for minimal $\eta$ for effective local linearity during sampling, as larger values of both $\gamma$ and $\eta$ amplifies the influence of secondary current peaks, thereby deteriorating waveform fitting. Therefore, a highly asymmetric gate pulse shape with low $\gamma$ is deemed optimal, necessitating $\eta$ to be as diminutive as possible. Our experimental validation illustrates that the $\gamma$ value of the THz waveform at 15\% humidity approximated 0.8, suggesting that an inner product angle less than 10º could be achieved with reasonably small $\eta$. Furthermore, a decrease in the $\gamma$ value from 0.8 to 0.2 incurs only a marginal reduction in the inner product angle, less than 1º, attesting that the $\gamma$ value in our experiment is sufficiently low for precise TDFES measurements. Supplementary insights in Fig. S2j reveals that the contribution of secondary pulse peaks to the transient THz current would be amplified at excessively high THz bias, imposing an upper limit on the Gate pulse strength. Additionally, the calculation of the full width at half maximum (FWHM) of the current pulse under varying THz electric field strengths, presented in Fig. S2i, indicates a linear increase in FWHM with THz maximum $E_{pk}$. Nonetheless, it is noteworthy that even with THz maximum $E_{pk}$ values as high as +700 V/cm, the resulting pulse width consistently remains below 200 femtoseconds. Furthermore, considering the 2 THz bandwidth of the THz pulse emitted from the LiNbO3 crystal\cite{RN2}, as illustrated in Fig. S1, and the narrowing effect due to tip antenna characteristics\cite{RN7,RN10,RN11}, the 200-fs-width gate pulse emerges as adequately sufficient for sampling.

\begin{figure}
  \centering
\includegraphics[width=0.9\textwidth]{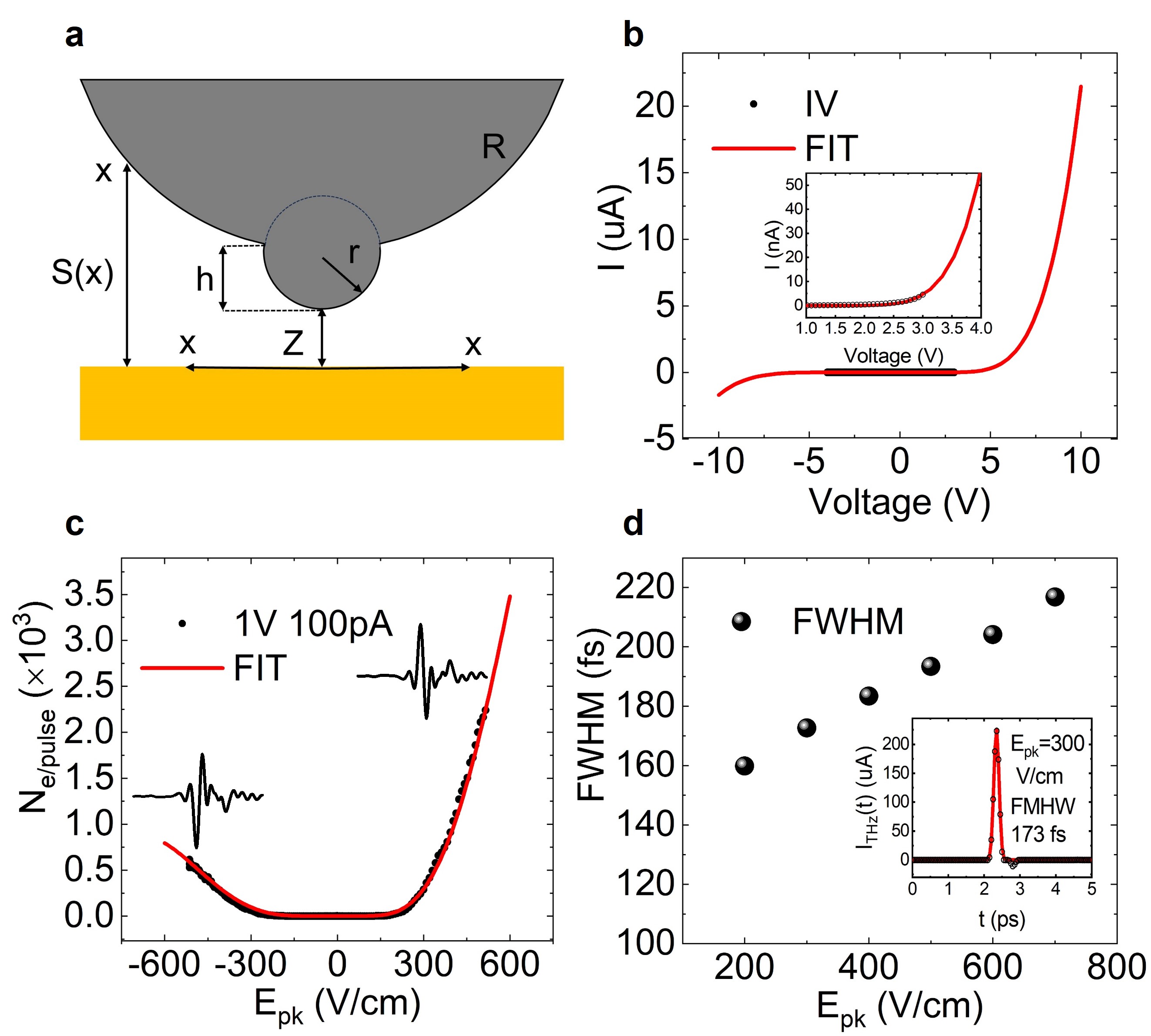}
  \caption{\textbf{Fitting with the Simmons model a. } The average barrier height (ABH) extracted from the current-tip height (I-Z) curves, recorded over a voltage range of -5 V to 5 V at 1 V intervals. \textbf{b}. The 3D model representation of the tip employed in the simulations. \textbf{c}. The curve illustrating the number of electrons per THz pulse (N) plotted against the THz maximum electric field peak ($E_{pk}$), complemented by its fitting curve. The negative section of the curve was acquired by reversing the carrier envelope phase of the THz pulse. This measurement was conducted at constant height mode with an STM setpoint of 1 V 100 pA (without THz illumination), and a DC bias of 0 V. \textbf{d}. The static current-voltage (IV) curve and its corresponding fitting curve measured at the same tip height as in (c), with an inset graph offering a magnified view. \textbf{e}. Far-field THz waveform shapes obtained at two distinct humidity levels and their corresponding current shapes predicted by the model. \textbf{f}. The simulated Gate pulse shape and the anticipated current pulse with a maximum positive $E_{pk}$ of +300 V/cm. The shape of the Gate pulse was characterized by the ratio of the negative maximum $E_{pk}$ to the positive maximum $E_{pk}$ ($\gamma=(|E_{pk-}|)/(|E_{pk+}|))$. \textbf{g}. Circles represent the target near-field (NF) waveform obtained from experimental TDFES results (at STM setpoint 3 V 20 pA, operating in constant height mode with a bias set to 3 V, Gate pulse maximum $E_{pk}$ of +463.7 V/cm, and a relative maximum $E_{pk}$ $\eta=E_{NFpk}/E_{Gatepk} of 0.18)$. Solid lines indicate simulated results with different $\gamma$values as in (f) (Gate pulse maximum $E_{pk}$ was +300 V/cm, and $\eta$ was 0.05). \textbf{h}. Inner product angles calculated between the target waveform and its fitting curve at various $\gamma$and $\eta$ values (Gate pulse maximum $E_{pk}$ was +300 V/cm). \textbf{i}. Gaussian fit Full Width at Half Maximum (FWHM) of the current pulse predicted by the model with varying Gate pulse maximum $E_{pk}$. \textbf{j}. Inner product angles calculated between the target waveform and its fitting curve at various $\gamma$and $\eta$ values (Gate pulse maximum $E_{pk}$ was +700 V/cm).}
  \label{fgr:example}
\end{figure}

\subsection{PES results at different illumination situations}

   \textbf{figure S3a} illustrates the photoemission sampling (PES) experiment, where an intense 515 nm laser pulse was focused on the STM junction, resulting in the emission of numerous electrons from either the sample or the tip\cite{RN4}. The voltage sensitivity of the photoemission current allowed for the acquisition of the THz near-field waveform by superposing the THz pulse with the 515 nm laser pulse. However, it is important to note that photoemission current can be emitted from both the tip and the sample sides. To clarify the PES results obtained under different illumination conditions, three scenarios were considered and depicted in \textbf{figure S3b}. In the case of sample emission, the current is high with a negative sample voltage, while no tunneling is observed with a positive sample voltage. This situation is completely reversed in the tip emission case, and there is an intermediate state where both the tip and the sample are illuminated. As a result, THz near-field waveforms were measured under both sample and tip illumination conditions and compared with the TDFES result. \textbf{figure S3c} shows that the near-field waveform in both conditions exhibits minor differences, except for an inverted carrier-envelope phase (CEP). Therefore, it is concluded that either tip emission or sample emission yields similar PES results when illuminating the Au (111) surface with a tungsten tip. Due to the smooth current curve obtained in the sample emission case, all PES experiments in the manuscript were conducted with sample emission current. A typical THz near-field waveform using sample photoemission current is displayed in\textbf{ figure S3d}, confirming the absence of other interfering THz pulses in the TDFES measurements.

\begin{figure}
  \centering
\includegraphics[width=0.8\textwidth]{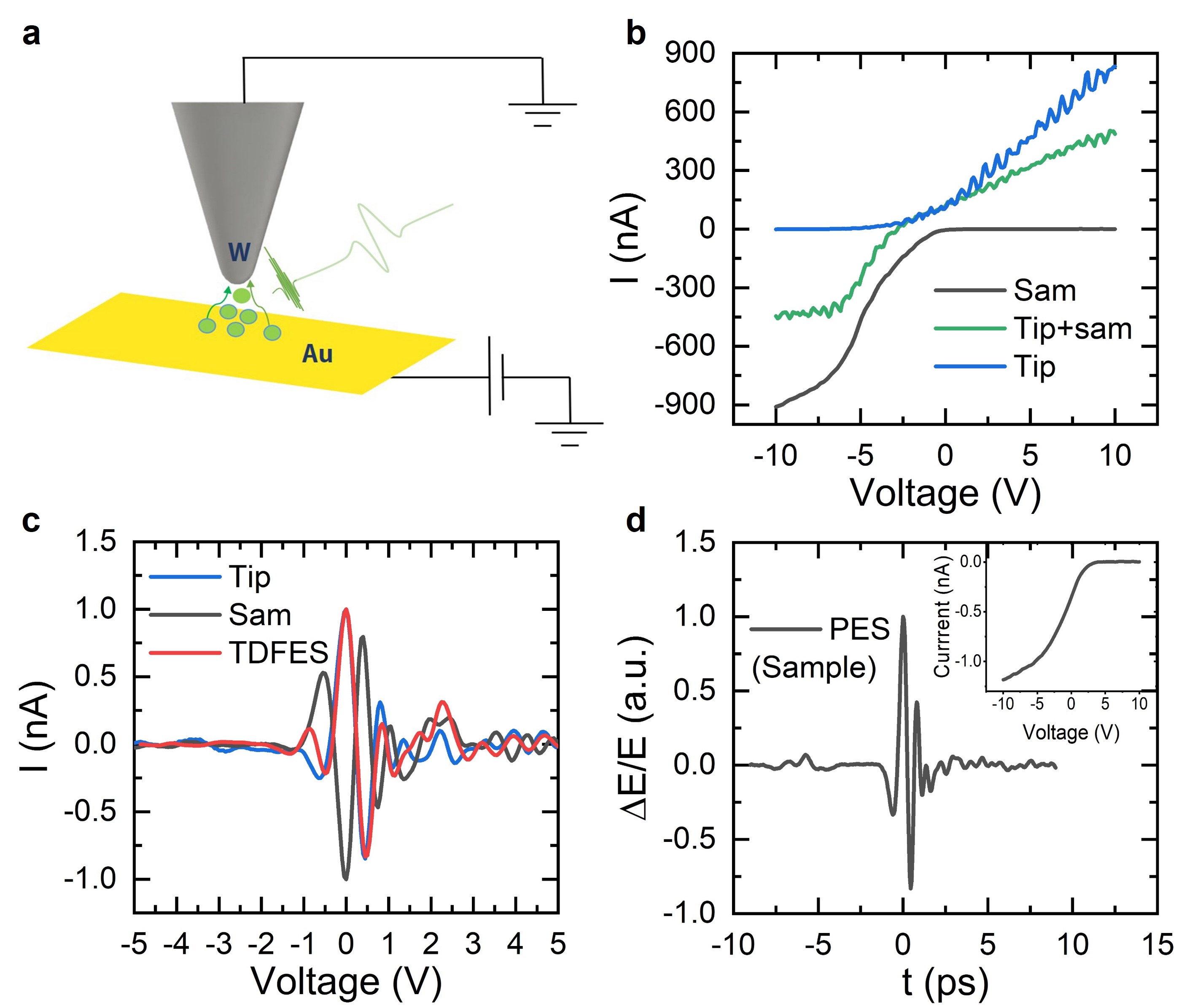}
  \caption{\textbf{Illustrations of the PES method} \textbf{a.} The tunneling picture of the photoemission process at the sample illumination situation. Positive voltage was applied and the green circles reflected the flow of the photoemission electrons. \textbf{b.} The photoemission current curves measured under different illumination conditions when illuminating the sample only, the tip only, or both the tip and sample, respectively. The pulse energy was 0.2 $\mu$J. \textbf{c. }The PES measured THz near-field waveforms under three illuminating conditions using the same tip. \textbf{d. }The PES result of a single THz pulse on the Au (111) surface under sample illumination condition. The bias voltage was -10 V, the current was -1.5 nA and the pulse energy was 0.6 $\mu$J. The inset chart shows the emission current curve utilized here and used in\textbf{ figure 3}.
}
  \label{fgr:example2}
\end{figure}

\subsection{THz electric field strength calibration}
The calculation of the THz electric field was conducted through the electro-optical sampling (EOS) method, relying on the detection of electrical signals A and B generated by two probe lights after passing through a Wollaston prism. The formula used for this calculation was:
\begin{equation}
    E_{THz-pk}=\frac{1}{t_{ZnTe}}(\frac{A-B}{A+B})(\frac{\lambda}{2\pi n_0^3r_{41}L}) \tag{7}
\end{equation}
Here, $\lambda$ represents the center wavelength of the probe light at 1030 nm, \textit{n\textsubscript{o}} = 2.7889 is the optical refractive index of ZnTe (110) crystal at 1030 nm, \textit{r\textsubscript{41}} = 4.45×10\textsuperscript{-12} m/V is the electro-optic coefficient for ZnTe\cite{RN9}, \textit{L} = 1 mm is the thickness of the ZnTe crystal, and \textit{t\textsubscript{ZnTe}} = 0.48 is the Fresnel amplitude transmission coefficient for a THz pulse passing through the ZnTe\cite{RN12}. This result was further calibrated to account for experimental conditions, considering factors including the focusing differences between the ZnTe crystal and the STM tip($\frac{f_{ZnTe}}{f_{STM}}=\frac{50.8}{33.85}=1.5007$), the power transmission rates of the diamond window (0.72) and two sapphire windows (0.52). The maximum peak intensity of the far-field THz electric field at the STM tip is then accurately determined by:
\begin{equation}
    E_{pk-tip}=\frac{1}{t_{ZnTe}}(\frac{A-B}{A+B})(\frac{\lambda}{2\pi n_0^3r_{41}L})(\frac{f_{ZnTe}}{f_{STM}})(\sqrt{T_dT_s}) \tag{8}
\end{equation}

\subsection{Calibration of the tunnel current from the lock-in current amplifier output}
    The signal processing procedures for calibrating the lock-in current output involved multiple stages, as illustrated in figure S4. Initially, the continuous THz pulse stream was modulated using a square wave reference signal and subsequently fed into a current pre-amplifier characterized by specific amplification parameters (10\textsuperscript{9} V/nA ) and a constrained low-pass bandwidth (10\textsuperscript{3} Hz). The resultant signal from the current amplifier was divided into two branches: the primary voltage signal was directed towards a lock-in amplifier (10/11), while the total current signal was transmitted to the Nanonis software for loop control. Inside the lock-in amplifier, the input signal underwent further signal conditioning, including high-pass filtering and multiplication with a coherent sinusoidal signal generated from the reference channel. The product of this multiplication was then filtered using a low-pass filter, resulting in the demodulated signal,\textit{ Demod X}, signifying the final output. 

Mathematically, this process can be expressed as following calculations. Firstly, the periodical THz tunneling current signal, \textit{g(t)}, with a repetition rate$f_r >> 1 kHz$, is expressed as: 

\begin{equation}
    g(t)=\sum_n f(t-nT_r), T_r=1/f_r \tag{9}
\end{equation}
where\textit{ f(t)} is the current signal of a single THz pusle, \textit{T\textsubscript{r}} is the repetition period of the THz pusle stream.Subsequently, this periodical signal was chopped at a frequency of$f_c < 1 kHz$, yielding a signal in the frequency domain as:
\begin{align}
    G(f)=\sum_na_n\delta(f-nf_r)\bigotimes\sum_nb_n\delta(f-nf_c)\notag\\
    =\sum_{m,n}a_nb_m\delta(f-nf_r-mf_c)\tag{10}
\end{align}
where $\bigotimes$ denotes the convolutional computation and the coefficients are:
\begin{equation}
    a_n=\frac{1}{T_r}\int_0^{T_r} f(t)exp(-j\cdot n2\pi f_rt)dt\tag{11}
\end{equation}
\begin{equation}
    b_n=\frac{1}{T_c}\int_0^{T_c} h(t)exp(-j\cdot m2\pi f_ct)dt=\frac{1}{2}sinc(\frac{\pi}{2}\cdot m)\tag{12}
\end{equation}

Here, \textit{h(t)} repesents the modulation function under the chopping condition characterized by a square-wave signal. The THz component of the output current from the DLPCA-200, \textit{I(t)}, can be expressed as:

\begin{equation}
    I(t)=I_0(\frac{1}{2}+\frac{2}{\pi}cos(2\pi f_ct)), I_0=a_0=\frac{1}{T_r}\int_0^{T_r}f(t)dt\tag{13}
\end{equation}
Thus, \textit{Demod X} is directly related to \textit{I\textsubscript{0}} as

\begin{equation}
    Demod X=\frac{10}{11\pi}I_0\tag{14}
\end{equation}
The number of electrons per THz pulse, \textit{N}, can be calculated from \textit{Demod X} using the formula:

\begin{equation}
    N=\frac{11\pi}{10f_re}\cdot Demod X\tag{15}
\end{equation}
where\textit{ e} represents the elementary charge and \textit{f\textsubscript{r}} is the repetition rate of the THz pulse stream. Equation (11) provides a rigorous calibration approach for the lock-in current amplifier output (\textit{Demod X}) in terms of electron count per THz pulse (\textit{N}).

\begin{figure}
    \centering
    \includegraphics[width=1\textwidth]{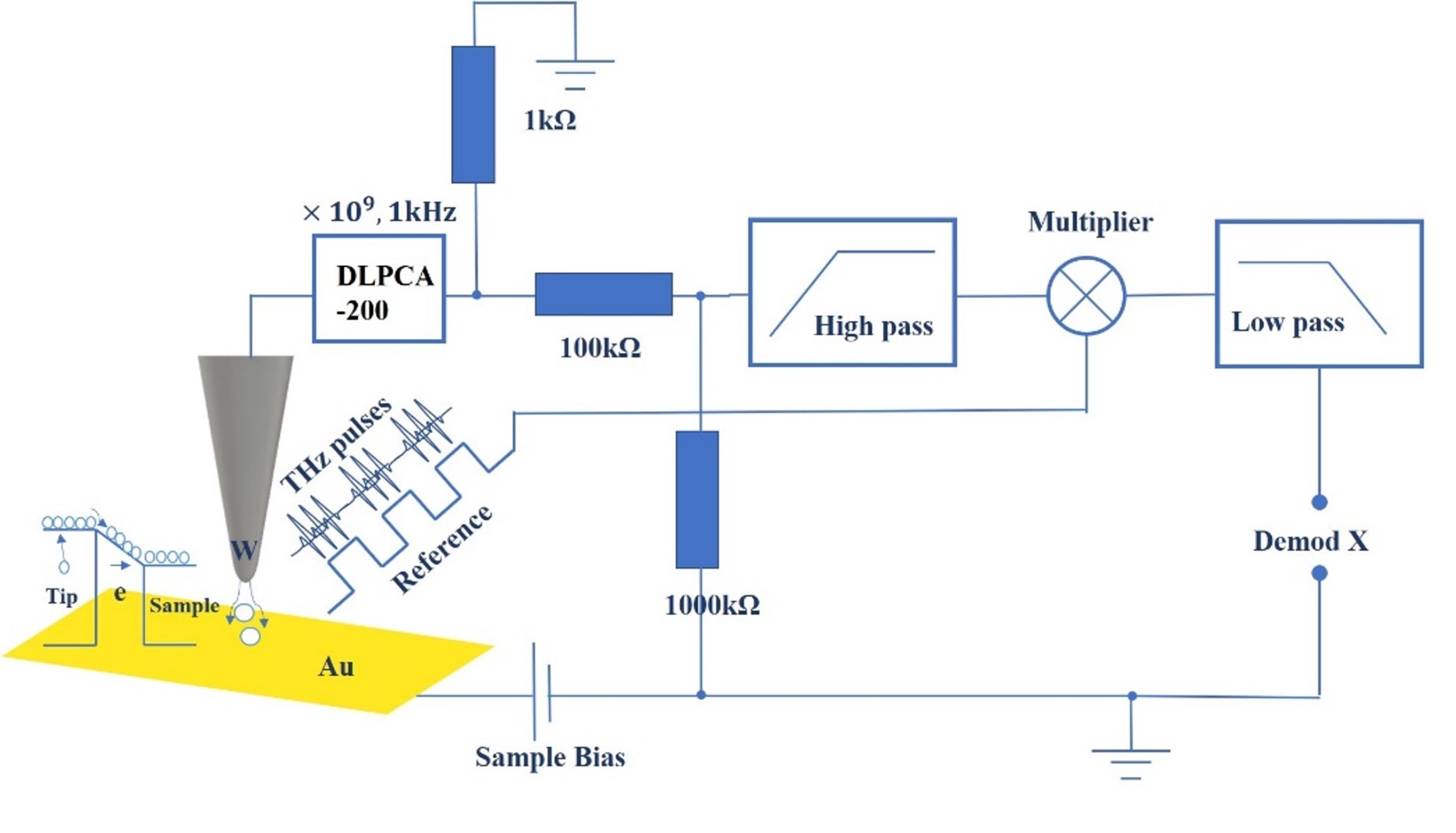}
    \caption{\textbf{Current calibration of the lock-in current amplifier output.} }
    \label{fig:enter-label4}
\end{figure}
\section{Other Supplementary Figures}
\begin{figure}
    \centering
    \includegraphics[width=1\textwidth]{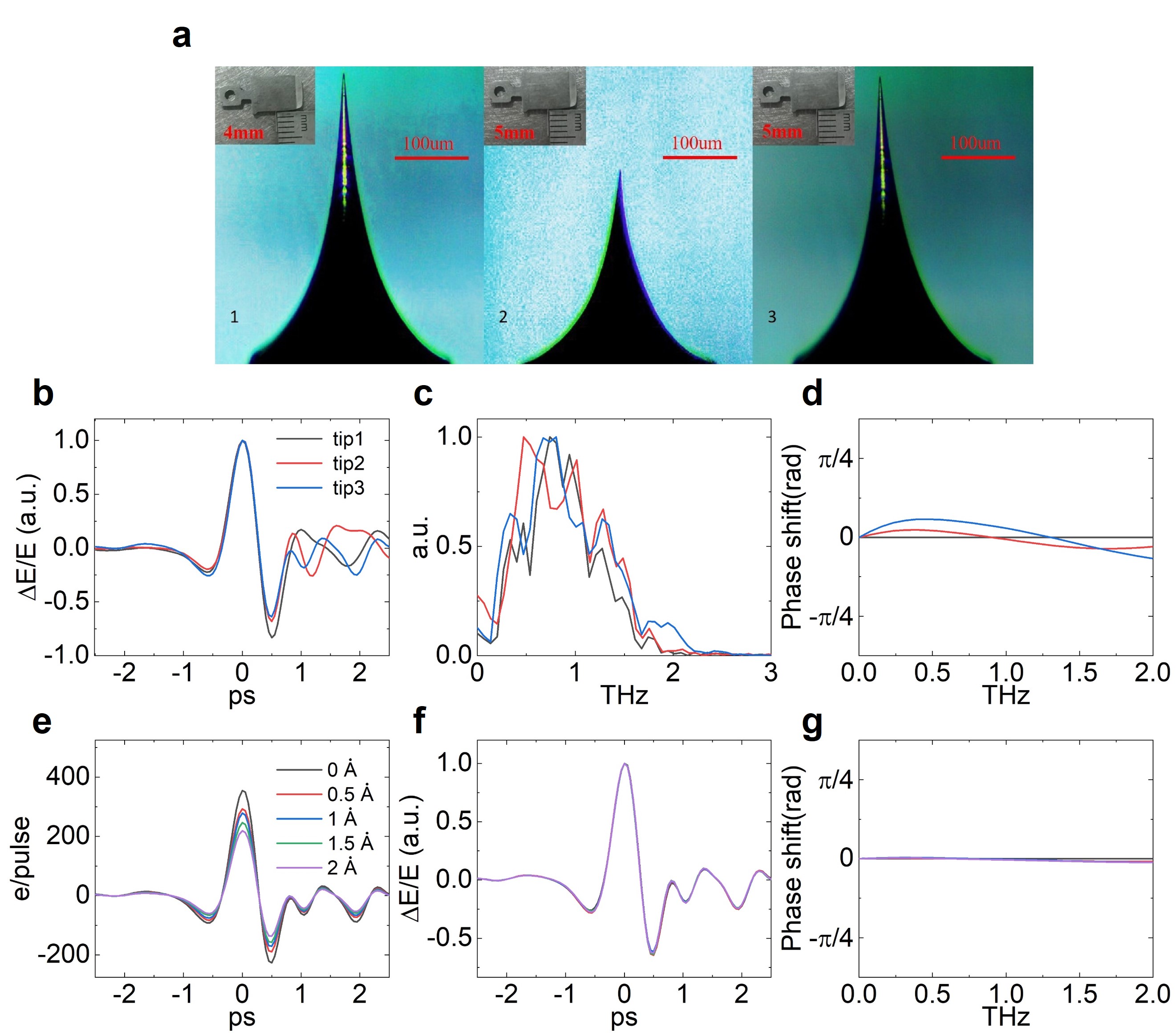}
    \caption{\textbf{Influence of the tip height and the tip configuration} \textbf{a. }Three tungsten tips with different shapes and lengths examined in experiment: Tip 1 (sharp, 4mm length), Tip 2 (blunt, 5mm length), and Tip 3 (sharp, 5mm length). \textbf{b.} The corresponding TDFES results in the time domain for these three tips, with data normalized to the maximum value. \textbf{c.} The FFT-transformed results. \textbf{d.} CEP shifts relative to Tip 1 for the three tips, indicating variations in the phase characteristics.\textbf{ e.} The impact of tip height on TDFES measurements recorded using Tip 3, where 0 Å refers to the tip height set at 3 V, 20 pA, and other heights were measured relative to this value. Tunneling electrons per THz pulse were calculated based on the lock-in current amplifier output, as detailed in \textbf{figure S4}. \textbf{f.} The corresponding time domain results with data normalized to the maximum value. \textbf{g.} The CEP shifts relative to 0 $\dot{A}$. In \textbf{(b)-(d)}, all the measurements were performed on a clean Au (111) surface under a tip height setting of 3 V, 20 pA (bias set to 3 V, constant height mode), at a set point of 664.4 V/cm, with relative intensities 0.18, 0.17, and 0.12 for the three tips respectively. In \textbf{(e)-(g)}, TDFES measurements were conducted under similar conditions in \textbf{(b)-(d)} except that Tip 3 was in use and the relative intensity was 0.12.} 
    \label{fig:enter-label5}
\end{figure}
\begin{figure}
    \centering
    \includegraphics[width=1\textwidth]{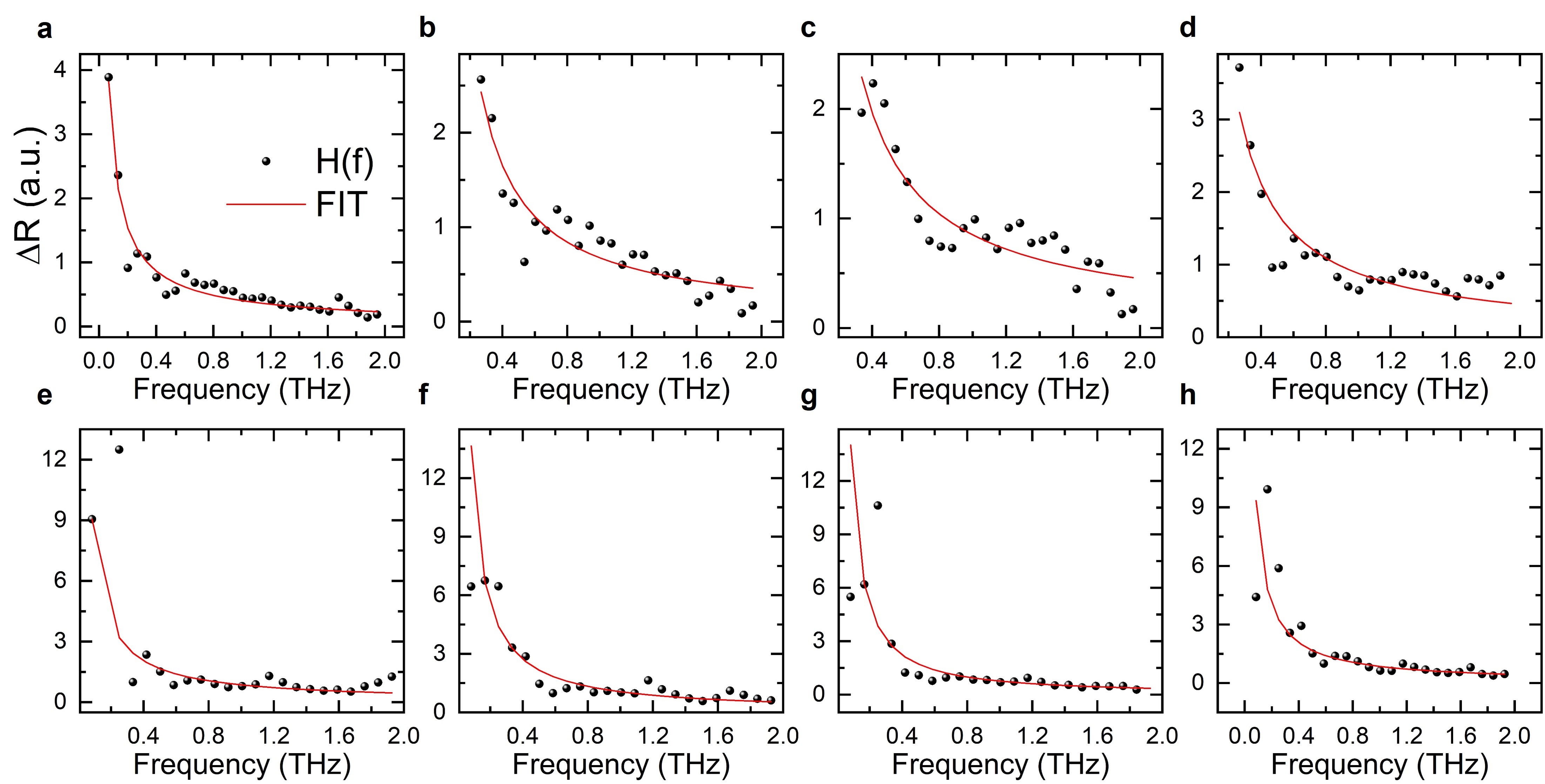}
    \caption{\textbf{Verification of the 1/f law in TDFES measurements} \textbf{a.} The transfer function measured with the tip used in \textbf{figure 2}, and the fitting curve follows a power-law form of $\sim f^{-0.83}$ \textbf{b-d.} The transfer functions measured with different tips from \textbf{figure S5}: \textbf{(b) }using Tip 1, with a fitting curve $\sim f^{-0.97}$; \textbf{(c) }using Tip 2, with a fitting curve $\sim f^{-0.91}$; \textbf{(d)} using Tip 3, with a fitting curve $\sim f^{-0.96}$. \textbf{e-h. }The transfer functions calculated from CEP-varied TDFES results in \textbf{figure 3}: \textbf{(e) }with a CEP shift of $-\pi /2$ and a fitting curve $\sim f^{-0.95}$; \textbf{(f) }with a CEP shift of $0$ and a fitting curve $\sim f^{-0.1.03}$; \textbf{(g)} with a CEP shift of $\pi /2$ and a fitting curve $\sim f^{-1.18}$; and \textbf{(h) }with a CEP shift of $\pi$ along with a fitting curve $\sim f^{-0.96}$}
    \label{fig:enter-label6}
\end{figure}
\begin{figure}
    \centering
    \includegraphics[width=1\textwidth]{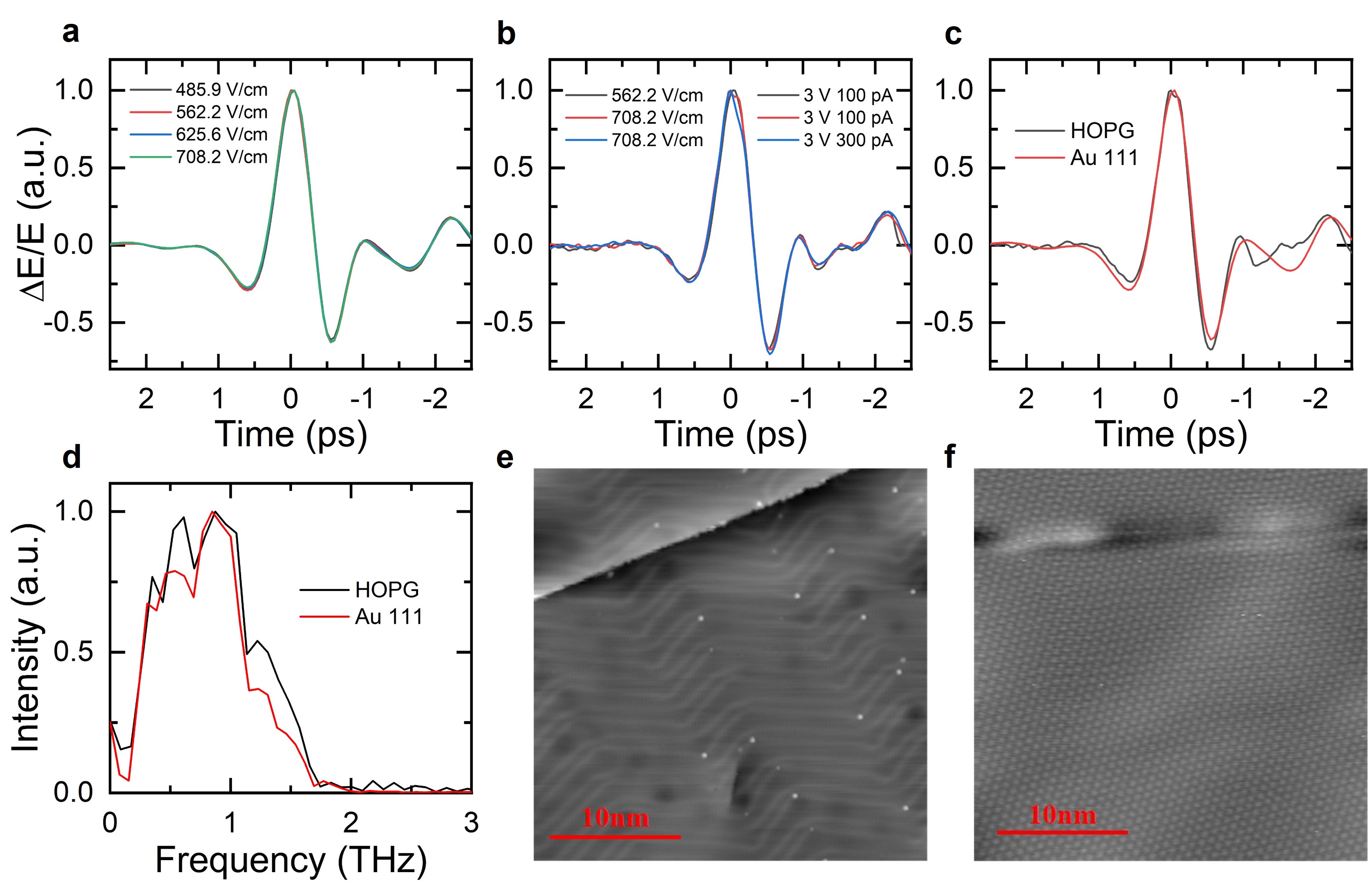}
    \caption{\textbf{TDFES results on Au (111) and HOPG surfaces}. \textbf{a. }The TDFES results on the Au (111) surface while varying the THz setpoint at: 485.9 V/cm, 562.2 V/cm, 625.6 V/cm, and 708.2 V/cm. These measurements were conducted with a tip height set at 3 V, 20 pA, and a relative intensity of 0.12. The convergence of results with increasing set point values demonstrates the correctness of the waveform. \textbf{b. }The TDFES results on highly oriented pyrolytic graphite (HOPG) while varying the sampling conditions, maintaining a relative intensity of 0.12. Increasing the THz setpoint and lowering the tip height do not alter the waveform, supporting the correctness of the waveform. \textbf{c-d.} The comparison of the TDFES results on Au (111) and HOPG surfaces in both the time domain \textbf{(c)} and the frequency domain \textbf{(d)}. \textbf{e. }The surface images of Au (111) obtained at the constant current mode. The STM bias setting was -1 V, 100 pA. \textbf{f. }The surface images on HOPG, with a STM bias setting of -1.5 V, 100 pA.
 }
    \label{fig:enter-label7}
\end{figure}
\clearpage
\bibliography{SI}